\documentclass[twocolumn,showpacs,preprintnumbers,amsmath,amssymb]{revtex4}

\usepackage{graphicx}
\usepackage{dcolumn}
\usepackage{bm}

\begin{document}


\title{``Pair'' Fermi contour and \\ repulsion-induced superconductivity in cuprates}

\author{V.I.Belyavsky}
\altaffiliation{State Pedagogical University, Voronezh, 394043,
Russia}

\author{Yu.V.Kopaev}
\affiliation{Lebedev Physical Institute, Russian Academy of
Sciences, Moscow, 117924, Russia}%

\date{\today}

\begin{abstract}
The pairing of charge carriers with large pair momentum is
considered in connection with high-temperature superconductivity
of cuprate compounds. The possibility of pairing arises due to
some essential features of quasi-two-dimensional electronic
structure of cuprates: (i) {\it The Fermi contour with strong
nesting features}; (ii)  {\it The presence of extended saddle
point near the Fermi level}; (iii) {\it The existence of some
ordered state (for example, antiferromagnetic) close to the
superconducting one as a reason for an appearing of ``pair'' Fermi
contour resulting from carrier redistribution in momentum space}.
In an extended vicinity of the saddle point, momentum space has
hyperbolic (pseudoeuclidean) metrics, therefore, the principal
values of two-dimensional reciprocal reduced effective mass tensor
have unlike signs. At small momenta of the relative motion of a
pair with large pair momentum, the pairing is sensitive just to
sign and value of the effective mass but not to only the value of
the Fermi velocity as it is in the case of Cooper pairing. Nesting
of the Fermi contour results in an increase of the statistical
weight of the pair with large total momentum due to an extension
of momentum space domain which corresponds to permissible values
of the relative motion momentum. Rearrangement of holes in
momentum space results in a rise of ``pair'' Fermi contour which
may be defined as zero-energy line for relative motion of the
pair. The superconducting gap arises just on this line. Pair Fermi
contour formation inside the region of momentum space with
hyperbolic metrics results in not only superconducting pairing but
in a rise of quasi-stationary state in the relative motion of the
pair. Such a state has rather small decay and may be related to
the pseudogap regime of underdoped cuprates. Bounded states of the
relative motion of the pair is studied both for attraction and
repulsion between the components of the pair. It is concluded that
the pairing in cuprates may be due to screened Coulomb repulsion.
In this case, the superconducting energy gap in hole-doped
cuprates exists in the region of hole concentration which is
bounded both above and below. The superconducting state with
positive condensation energy exists in more narrow range of doping
level inside this region. Such hole concentration dependence
correlates with typical phase diagram of cuprates. The pairing
mechanism and the pair Fermi contour conception make possible a
rise of the superconducting condensate and quasi-stationary states
of pairs and may provide qualitative interpretation for the key
experimental facts relating to cuprates, namely: (1) Observable
values of the superconducting transition temperature; (2) The same
symmetry and the same energy scale of the superconducting gap and
the pseudogap; (3) Relatively small values of the coherence
length; (4) Asymmetry of tunnel current-bias characteristic; (5)
An ``apparent'' violation of the Ferrell-Glover-Tinkham
low-frequency optical sum rule; (6) Quasi-particle peak and
``dip-hump'' structure of the angle-resolved photoemission and
tunnel spectra; (7) ) Anomalous large (as compared with the
prediction of the theories of the Bardeen-Cooper-Schrieffer type)
values of $2\Delta/T_C$ ratio observed in extremely underdoped
cuprates with hole doping; (8) The pair Fermi contour conception
does not require any additional suggestion relating to a character
of carrier or pair scattering such as the so-called ``forward
scattering'' or ``hot and cold spots'' on the Fermi surface; (9)
The pair Fermi contour conception is fairly consistent with a
character of doping dependence of pseudogap state crossover
temperature and also both superconducting transition temperature
and superfluid density observed in HTSC cuprates.
\end{abstract}

\maketitle

\section{ Introduction}

Recently, we proposed new mechanism of superconducting (SC)
pairing in anisotropic quasi-two-dimensional (2D) electron system
typical of high-temperature superconducting (HTSC) cuprate
compounds \cite{1, 2, 3, 4, 5, 6, 7}. Pairs with large momentum
${\bm K}$ (${\bm K}$--pairs) are considered; here, $K\approx
2{k}_F$, $k_F$ is the value of the Fermi momentum directed along
${\bm K}$. It is well known \cite{8} that the Cooper channel of
pairing becomes inefficient when the pair momentum exceeds a value
of the order of $\Delta /{v}_F$; here, $\Delta$ is the SC gap at
$K=0$, $v_F$ is the Fermi velocity. The same relatively small
value of the pair momentum corresponds to the wavelength of
spatially inhomogeneous SC phase arising in weakly ferromagnetic
electron system \cite{9,10} as well. The Cooper channel at $K \neq
0$ is suppressed due to Pauli's exclusive principle which
restricts the phase volume accessible for the electron states
contributing to the ${\bm K}$--pair state. This phase volume
decreases rapidly with $K$ and vanishes at $K\sim {\Delta
/{v_F}}$. Therefore, pairing with large pair momentum may be
possible if some rearrangement in the electron system provides
finite (and sufficiently large) phase volume for the states
forming  ${\bm K}$--~pair. Any rearrangement of electrons in
momentum space which transfers a part of them across the Fermi
surface (FS) results in an increase of the energy of the electron
subsystem of a crystal. If the electron subsystem interacts with
some other one, for example, antiferromagnetic (AF) spin
subsystem, and such an interaction results in a gain in total
energy, new FS arises corresponding to new ground state of the
system. All of undoped HTSC cuprates are AF insulators, therefore,
the pairing mechanism \cite{1, 2, 3, 4, 5, 6, 7} is not exotic,
most likely, this mechanism is typical of doped HTSC cuprate
compounds. Electronic structure and physical properties of such
layered compounds as HTSC cuprates are studied in detail \cite{11,
12}. Conducting $CuO_2$ plane (one or more per unit cell) is the
key structural unit of any HTSC cuprate compound. The neighboring
planes are separated by reservoirs that is atomic layers which,
under doping, inject carriers (holes, as usual) into these planes.
The interplane coupling is very weak and this is just the reason
one may consider HTSC cuprates as 2D electron system. In undoped
cuprates, there is long-range AF order below the Neel temperature,
$T_N$. Doping leads to progressive destruction of long-range order
and, as a final result, to a rise of strongly anisotropic metallic
(M) state. Angle-resolved photoemission spectroscopy (ARPES)
measurements \cite{14, 15, 16, 17, 18} result in unambiguous
conclusion that, in the normal (N) state, any HTSC cuprate has
large FS. Observed FS's are in good agreement with band structure
calculations \cite{19, 20, 21, 22, 23} based on the density
functional theory. In the case of hole-doped compounds, the Fermi
contour (FC) that is the cross-section of the FS which is parallel
to conducting layers is a square with rounded corners. The FC of
holes is centered at $(\pi ,\pi)$ and exhibits strong nesting
feature along [100]-type directions. This is an evidence in behalf
of the significant role of the interaction between
next-nearest-neighbor atoms as far as taking account of
nearest-neighbor interaction only results in the square FC with
perfect nesting along [110]-type directions exactly at
half-filling \cite{11}. It should be noted that, in the case of
electron-doped compounds, such as $Nd_{2-x}Ce_{x}CuO_4$, any
appreciable nesting of the FC is absent and the FC is closed to a
circle. At approximately half-filling, long parts of the FC are
situated close to the saddle points of hole dispersion \cite{11}.
Hole doping moves the Fermi level towards the saddle point whereas
electron doping acts in reverse direction. Therefore, in
hole-doped compound, nesting feature of hole FC appears in
relatively wide concentration range. Weak dispersion along the
nesting directions results in the fact that longitudinal (along
the nested straight-line parts of the FC) component of the Fermi
velocity sufficiently smaller than the transversal one \cite{24}.
This corresponds to an effective enhancement of 2D density of
states in the vicinity of logarithmic van Hove singularity due to
the saddle point \cite{25}. Thus, there is an extended vicinity of
the saddle point in which the principal values of 2D tensor of
reversed reduced effective mass have unlike signs. One can say
that, in such a vicinity, momentum space has {\it hyperbolic}
(pseudoeuclidean) metrics. Due to nesting feature of the FC, the
absolute values of the principal effective masses differ strongly
from each other: positive longitudinal mass is essentially more
than the absolute value of negative transversal mass. It should be
pointed out once more that, in a case of any hole-doped cuprate
compound, long straight-line parts of the FC are situated, in
main, just in such ``{\it flat-band}'' or ``{\it extended van Hove
singularity}'' vicinity \cite{11}.

In HTSC cuprates, the SC state appears in some doping interval,
$x_{\ast}<x< x^{\ast}$, bounded both above and below. Both
superconducting transition temperature $T_C$ and superfluid
density (or phase stiffness) ${\rho}_s$ may demonstrate highly
complicated dependence on doping in this interval \cite{12,26}.
The absolute maximum of $T_C$ corresponds to the optimal doping,
$x_{opt}$. Phase diagram typical of hole-doped HTSC cuprates is
presented in Fig.~1.

In underdoped ($x < x_{opt}$) compounds, one-particle density of
states is suppressed essentially at $T_C < T < T^{\ast}$. Such a
suppression may be interpreted as a rise of the so-called
pseudogap in the excitation spectrum \cite{27}. The temperature
$T^{\ast}$ corresponding to a crossover between the N state at $T
> T^{\ast}$  and the ``pseudogap regime'' at $T_C <T < T^{\ast}$
decreases with doping increase and becomes approximately equal to
$T_C$ at $x \simeq x_{opt}$. The pseudogap ${\Delta}^{\ast}$, just
as the SC gap $\Delta$, is strongly anisotropic, and also, the
character of the anisotropy is the same both for ${\Delta}^{\ast}$
and $\Delta$ \cite{12}. The maxima of their absolute values
correspond to antinodal [100]-type directions. The minimal values
(which, possibly, are equal to zero) both of ${\Delta}^{\ast}$ and
$\Delta$ correspond to nodal [110]-type directions. Pseudogap
evolution with temperature decrease from $T^{\ast}$ to $T_C$ was
studied using ARPES technique \cite{28} in underdoped single
crystals $Bi_{2}Sr_{2}CaCu_{2}O_{8+x}$. The pseudogap arises at $T
= T^{\ast}$ at four points of the FC corresponding to antinodal
directions. Lowering of temperature from $T^{\ast}$ leads to
pseudogap extension in the directions of the corners of the FC.
Thus, the FC turns out to be discontinuous and has the form of
four arcs (rounded corners of the square FC) which are not
connected with each other \cite{25}. The arc length decreases
gradually with temperature lowering. At $T = T_C$, the FC
disappears and, instead of it, the SC gap arises which has minimal
(or equal to zero) values just in the points (corresponding to
nodal directions) where the FC shrinks \cite{28}. Knight shift
measurements indicate that there is singlet pairing of carriers
when the electron system of HTSC cuprates is in the SC state
\cite{29,30,30'}. Therefore, observed momentum dependence of the
SC gap may correspond to either anisotropic $s$-type or $d$-type
of orbital symmetry \cite{31}. The same orbital symmetry and the
same energy scale of $\Delta$ and ${\Delta}^{\ast}$ \cite{32}
enable one to suppose that the SC gap and the pseudogap are of the
same origin. Thus, the pseudogap regime may be considered as an
incoherent state of paired charge carriers \cite{33, 34}.

In the theory by Bardeen, Cooper and Schrieffer (BCS) which
explains the conventional superconductivity successfully
\cite{35}, attraction due to virtual phonon exchange is a driven
force leading to pairing of carriers. In principle, phonon
mechanism of Cooper pairing should not be excluded as a mechanism
of HTSC \cite{25} although it is difficult to explain
satisfactorily some essential features of HTSC state, for example,
the symmetry of the SC gap. In view of the fact that the phase
diagram of any HTSC cuprate has a region with long-range AF order,
AF fluctuation exchange as a mechanism of pairing \cite{36, 37,
38, 39} seems as quite natural (neutron scattering experiments
\cite{40, 41, 42, 43} exhibit broadened Bragg peaks up to the
optimal doping). The other point of view is founded on the
statement that ground state energy gain at the SC transition in
HTSC cuprates is due to a lowering of the kinetic energy arising
when two of like-charged carriers form a pair \cite{44, 45, 46,
47, 48}. In such a case, generally speaking, one needs no
attraction between carriers and screened Coulomb repulsion remains
as a natural essential interaction in the electron system.

AF fluctuations (short-range AF order) may lead to a specific
quasi-one-dimensional (1D) self -- organization in 2D electron
system of HTSC cuprates. Elastic neutron scattering study in
$La_{1.6-x}Nd_{0.4}Sr_{x}CuO_{4}$ compound enables one to assume
that, at $T < T_s \approx 80 K$, holes doped into a crystal are
situated in 1D antiphase boundaries (charge stripes) separating
hole depleted domains with AF order \cite{49, 50}. A rise of such
static stripe structure (in a general way, predicted in \cite{51,
52, 53}) may be described as a transfer of excess holes from AF
part of a stripe into antiphase boundary (M part of a stripe). In
underdoped $La_{2-x}Sr_{x}CuO_{4}$ which does not contain Nd
atoms, dynamic (fluctuating) stripes were observed \cite{54}.
Dynamic stripe magnitude, just as the magnitude of AF
fluctuations, decreases with doping and, at $x> x_{opt}$, neutron
scattering technique does not make possible a resolution of
strongly broadened stripe peaks of rather low intensity. A stripe
structure may exist being independent of superconductivity but
such a structure (just as AF fluctuations) and superconductivity
are closely and in a nontrivial way connected with each other. As
an indirect confirmation of this statement one may take into
consideration the fact that, in $La_{1.6-x}Nd_{0.4}Sr_xCuO_4$ at
$x = 1/8$ when static stripe magnitude is maximal, there is a
local minimum on the doping dependence of the SC transition
temperature \cite{55}]. On the contrary, it is possible that
dynamic stripes stimulate superconductivity \cite{12}.

Experimental data available make possible to determine the main
features and details of the electronic structure which are
essential to understand the character of the SC state of HTSC
cuprates and interpret their physical properties qualitatively.
Firstly, all doped HTSC compounds have 2D electronic structure
with strong nesting of the FC situated in an extended vicinity of
the saddle point of the hole dispersion. Secondly, in all doped
HTSC compounds, doping regions corresponding to AF and SC phases
are close to each other and, in the SC region, there is
short-range AF order resulting in stripe self-organization of spin
and charge subsystems of the crystal. The theory here developed
takes into account these principal features of the electronic
structure and can qualitatively explain the key experimental facts
relating both to N and SC state of HTSC cuprates.

The paper is organized as follows: The next section is dedicated
to the formulation of the conditions under which the pairing with
large pair momentum may be possible; in addition, we introduce
the concept of the ``pair'' Fermi contour. In Sec.~III we
consider the problem of a single pair in momentum space with
hyperbolic metrics and discuss the symmetry properties of the
pair wave function. Sec.~IV contains a discussion of the
character of the two poles of the scattering amplitude
corresponding to a quasi-stationary state of the pair and
superconducting instability. Effective interaction between
particles composing a pair with large momentum is considered in
Sec.~V. In Sec.~VI we derive the equation defining the SC order
parameter. The approximate solutions of this equation are
presented in Sec.~VII both in the case of attraction and
repulsion between the components of the pair. In Sec.~VIII, we
discuss SC transition induced chemical potential shift, and then,
in Sec.~IX, doping dependent SC condensation energy is studied.
In Sec.~X, we consider some special case of SC state arising due
to weak ferromagnetism, associated with stripe structure. Sec.~XI
is dedicated to a brief discussion of some key experimental
results related to both N and SC state of HTSC cuprates; also, we
discuss some other possible reasons of ``opening'' of the PFC and
propose a qualitative interpretation of available experimental
data in the scope of the theory developed here.


\section{ Electron and hole pairs. Pair Fermi contour}

Let us consider two electrons or two holes with total momentum
${\bm K}={{\bm k}_+}+{{\bm k}_-}$ where ${{\bm k}_+}$ and ${{\bm
k}_-}$ are momenta of the particles composing a pair. This is a
pair of noninteracting particles. Thus, now and below in this
Section, we do not fall outside the limits of usual one-particle
approximation and the pair here introduced may be named as a {\it
slave pair}. Further, taking account of the screened Coulomb
interaction between particles composing the pairs, we use such
pairs to construct a SC state. Filling of the states inside the FC
results in the fact that permissible values of a momentum of the
relative motion of the $\bm K$--pair, ${\bm k}=({{\bm k}_+}-{{\bm
k}_-})/2$, belong to a certain domain of momentum space. Such a
domain, which we denote as $\Xi_K$, has a form dependent on $\bf
K$ and on a shape of the FC \cite{1,2}. Typical of any hole-doped
HTSC cuprate FC is a square with rounded corners as it is
represented schematically in Fig.~2. If the pair momentum is
directed along [100] and $K<2k_F$ the corresponding domain has a
form shown in the same Figure. It is clear from Fig.2 how one can
define such a domain at any given $\bm K$. It should be noted that
the area (labeled by the same symbol $\Xi _K$) of the domain
$\Xi_K$ tends to zero when $K \to 2k_F$. Thus, the statistical
weight of the states which compose pairs with $K=2k_F$ is equal to
zero even in the case of perfect nesting \cite{56, 57}. It should
be noted that the model used in \cite{56, 57} does not take into
account the existence of an extended saddle point and hyperbolic
metrics of momentum space, therefore, a pairing with $K=2k_F$
turns out to be impossible. For arbitrary direction of $\bm K$,
there are, generally speaking, eight domains corresponding to the
pair momenta which are equivalent to given vector $\bm K$ (for
special, antinodal or nodal, direction there are four equivalent
vectors). However, at preassigned deviation of $\bm K$ from
$2k_F$, the statistical weight (the area ${\Xi }_K$) depends on
the $\bm K$ direction decreasing from a maximal value for
antinodal to a minimal one for nodal directions. It is quite
obvious that the binding energy of $\bm K$--pair has to increase
with ${\Xi }_K$, therefore, one may expect that a rise of the SC
condensate should be due to pairs with momenta corresponding to
antinodal directions.

There is an experimental evidence in behalf of the consideration
of hole pairs with large total momentum. As an example, one may
consider the so-called ``commensurate''neutron resonance (41
meV) peak below $T_C$ which is usually associated with a rise of
resonance collective triplet $\pi$--mode \cite{57'}  corresponding to
the saddle point. Recently, in \cite{57'',57'''} was observed
``incommensurate'' magnetic fluctuations  in HTSC
cuprates . As it is pointed out in Ref.\cite{57'}, the
incommensurate mode transforms continuously into the commensurate
one demonstrating, in addition, a negative (downward away from the
commensurate momentum) dispersion  \cite{57''''}. Such a tendency
of softening of this triplet $\pi$--mode can be interpreted as an
indirect evidence of the phase transition possibility associated
with softening of a certain singlet mode corresponding to large and
``incommensurate'' pair momentum.

Almost straight-line parts of the FC belong to the region of
momentum space with hyperbolic metrics. Therefore, the energy of
the relative motion of a pair inside $\Xi_K$,
\begin{equation}
 \label{1}\varepsilon_r({\bm K}, {\bm k})=
 \varepsilon\left({{\bm K}\over 2}+{\bm k}\right)+
 \varepsilon\left({{\bm K}\over 2}-{\bm k}\right)
 -2\varepsilon\left({{\bm K}\over 2}\right),
 \end{equation}
at relatively small $\bm k$ may be approximately represented as
\begin{equation}
  \label{2}\varepsilon_r({\bm K}, {\bm k})
  \approx {\frac {{\hbar}^2}{2m}}({{\nu}k_1^2}-k_2^2),
\end{equation}
where $\varepsilon(\bm k)$ is hole dispersion and, as it follows
from a symmetry consideration, the coordinate axes are directed
parallel (the $k_1$--axis) and perpendicular (the $k_2$--axis) to
the FC (Fig.~2). These coordinate axis directions correspond to
the principal directions of 2D tensor of reduced reciprocal
effective mass ($\nu /m$ and $-1/m$ are dependent on $\bm K$
principal values of this tensor). Due to strong nesting of the FC,
the absolute values of the effective masses differ considerably
from each other, namely, a dimensionless parameter $\nu <<1$.

The domain $\Xi_K$ consists of two parts, $\Xi_K^{(-)}$ and
$\Xi_K^{(+)}$, in which the energy of the relative motion of $\bm
K$-- pair is negative and positive, respectively. The domain
$\Xi_{K^\prime}$, also shown in Fig.~2, corresponds to a pair with
total momentum  $\bf K^\prime$ ($\bf K^\prime$--pair) outside of
the FC; thus $K^\prime > 2k_F$. This domain consists of two parts,
$\Xi_{K^\prime}^{(-)}$ and $\Xi_{K^\prime}^{(+)}$, corresponding
to negative and positive energy of the relative motion of $\bm
K^\prime$--pair as well. Excitations composing $\bf K$--pair
inside the FC are electrons whereas holes are the excitations
which compose $\bm K^\prime$--pair outside of the FC. Both $\Xi_K$
and $\Xi_{K^\prime}$ belong to the region of momentum space inside
2D Brillouin zone which has hyperbolic metrics.

In contrast to Cooper pairs with $K=0$, the energy of the relative
motion of $\bm K$--pairs at $K\approx2k_F$ is sensitive not only
to the value of the Fermi velocity but to signs and absolute
values of the effective masses. In the case of hyperbolic metrics,
$\bm K$--pair density of states exhibits a logarithmic van Hove
singularity corresponding to zero energy of the relative motion as
it is shown schematically in Fig.~3. Weak dispersion along one of
the directions in 2D momentum space (the $k_1$--axis in Fig.~2)
leads to the fact that $\bm K$--pair density of states has almost
1D character \cite{25}.

By definition of the ground state of the electron system, all pair
states inside $\Xi_K$ are occupied whereas the states inside
$\Xi_{K^\prime}$ are vacant. Such a filling of the states in
momentum space corresponds to spatially homogeneous state of the
electron system. At given $\bm K$, states of the relative motion
of $\bm K$--pair are characterized by the relative motion density
of states, ${g_K}(\varepsilon)$. Upper edge of pair density of
states in the domain $\Xi_K$ corresponds to the Fermi level
(Fig.~3). There is a finite energy gap $\delta
\varepsilon_{KK^\prime}$ between the upper edge of
${g_K}(\varepsilon)$ and the lower one relating to pair density of
states, ${g_{K^\prime}}(\varepsilon)$, in the domain
$\Xi_{K^\prime}$ as it is seen from Fig.~3. Therefore, any
transfer of a pair from $\Xi_K$ into $\Xi_{K^\prime}$ is
necessarily connected with an energy increase due to an increase
of center-of-mass energy. However, it should be noted that the
pairs having positive energy of the relative motion leave the
domain $\Xi_K$ whereas the pairs with negative energy arrive at
the domain $\Xi_{K^\prime}$ .

Such transfers of pairs from the domain $\Xi_K$ into a region of
momentum space outside of the FC may bear a relation to well-known
spatially inhomogeneous (stripe) structure in which there is an
alternation of hole enriched and depleted 1D regions \cite{12}.
The region of momentum space into which $\bm K$--pairs may
transfer is either that part, ${\tilde \Xi}_K$, of the domain
$\Xi_K$ which is situated outside of the FC or the domain
$\Xi_{K^\prime}$ corresponding to total pair momentum $\bm
K^\prime$ (Fig.~2). As far as relative motion density of states
which corresponds to $\bm K^\prime$--pair belonging to the
subdomain $\Xi_{K^\prime}^{(-)}$  is considerably greater than
density of states corresponding to the subdomain ${\tilde \Xi}_K$
we may restrict ourselves to a consideration of the transfers
$\Xi_K^{(+)} \rightarrow \Xi_{K^\prime}^{(-)}$ only. Suppose that
a number of pairs, $\delta N$, passes from $\Xi_K^{(+)}$ into
$\Xi_{K^\prime}^{(-)}$, so that in the subdomain $\Xi_K^{(+)}$ (in
the zero-temperature limit and in the absence of an interaction
between holes), vacant pair states arise in a certain (small in
comparison with $\delta \varepsilon_{KK^\prime}$) energy interval
near $2E_F$. The same number of pairs occupies a small energy
interval (which may be determined using pair number conservation
condition) near the lower edge of the band corresponding to the
subdomain $\Xi_{K^\prime}^{(-)}$. Thus, the energy increase due to
pair transfers from $\Xi_K^{(+)}$ into $\Xi_{K^\prime}^{(-)}$ may
be estimated as $\delta N \cdot \delta \varepsilon_{KK{^\prime}}$.

Transfers of pairs, $\Xi_K^{(+)} \Rightarrow
\Xi_{K^\prime}^{(-)}$, in momentum space may be related to
transfers of holes from AF parts of stripes into M parts in real
space (Fig.~4). An enhancement of AF correlations due to such
transfers results in some reducing of the energy which might
compensate the energy increase due to the excitation of hole pairs
leading to transfers $\Xi_K^{(+)} \Rightarrow
\Xi_{K^\prime}^{(-)}$. An energy gain due to a removal of $\bm
K$--pairs from $\Xi_K^{(+)}$, that is from AF parts of stripes,
may be estimated phenomenologically if one introduces depending on
doping parameter $I=I(x)$ which may be treated as nearest-neighbor
spin correlation function being a measure of AF short-range order.
Let us assume that each hole pair transferring from the subdomain
$\Xi_K^{(+)}$ into the subdomain $\Xi_{K^\prime}^{(-)}$ gives an
energy gain equal to $I$. Then, total decrease in the energy of
holes due to such transfers of $\delta N$ hole pairs may be
estimated as $-\delta N \cdot I$. Thus, a rise of the stripe
structure lowers the ground state energy provided that
\begin{equation}
\label{3}I > \delta \varepsilon_{KK^\prime}.
\end{equation}
The existence in the domain $\Xi_K =\Xi_K^{(-)}+\Xi_K^{(+)}$ of
hole filled part (for which we use above introduced notation
$\Xi_K^{(-)}$) and vacant part ($\Xi_K^{(+)}$) makes possible
pairing of carriers in the vicinity of the lines separating filled
and vacant subdomains. The energy of the relative motion of $\bm
K$--pair with respect to the value of the chemical potential is
negative inside $\Xi_K^{(-)}$  and positive inside $\Xi_K^{(+)}$,
therefore, the lines separating these subdomains (the lines of
zero relative-motion energy) play role of a peculiar ``pair''
Fermi contour (PFC) on which the SC gap may arise. Such a
conclusion is related both to the domains $\Xi_K$ and
$\Xi_K^\prime$, therefore, PFC is situated both inside and outside
of the parent FC. If the value of the vector ${\bm K}-{\bm
K^\prime}$ which may be considered as a reciprocal spatial scale
of the stripe structure appreciably exceeds a character scale,
${\delta k_c}\sim \Delta$, of non-zero SC order parameter in
momentum space, one may consider the pairing problems in $\Xi_K$
and $\Xi_{K^\prime}$ independently from each other. In the
following, we consider just the case when $|{\bm K}-{\bm
K^\prime}|>>\delta k_c$.


\section{ Problem of a single pair}

First of all, let us consider a hole $\bm K$--pair taking into
account two-particle potential interaction between the particles
composing the pair. As stated above, the momenta of interacting
particles are confined inside the domain $\Xi_K$. We suppose that
this domain belongs to a region of momentum space with hyperbolic
metrics. A wave function of $\bm K$--pair may be written as
\begin{equation}
\label{4}{\Psi_K}({\bm r_+},{\bm r_-})={\frac 1{\sqrt {S}}}
{\varphi_K} ({\bm r}) {e^{i{\bm KR}}}.
\end{equation}
Here, $\bm r_+$ and $\bm r_-$ are radius vectors of the particles,
${\bm R}=({\bm r_+}+{\bm r_-})/2$, ${\bm r}={\bm r_+}-{\bm r_-}$,
${\varphi_K}(\bm r)$ is a wave function of the relative motion,
$S$ is a normalizing area.

On account of the crystal symmetry, all wave functions
$\varphi_{{\hat g}K}$ corresponding to the momenta ${\hat g}\bm K$
turn out to be equivalent; here, $\hat g$ is a crystal symmetry
group transformation. Therefore, the $\bm K$--pair wave function
taking into account the crystal symmetry should be represented as
a linear combination of the form
\begin{equation}
\label{5} \Psi_K^{(\Gamma )} = \sum_{[{\hat g}K]} c_{{\hat g}K}
^{(\Gamma )} {\Psi_{{\hat g}K}}.
\end{equation}
A choice of the coefficients $c_{{\hat g}K}^{(\Gamma )}$ is
determined by the irreducible representation $\Gamma $ of the
crystal symmetry group according to which the wave function
Eq.~(5) transforms under the action of crystal symmetry operators
$\hat g$. It should be noted especially that the wave function
Eq.~(5) corresponds to current-less state in view of the fact that
$\sum {{\hat g}\bm K}=0$.

Taking account of the fact that the domains $\Xi_{{\hat g}K}$
corresponding to equivalent momenta, ${\hat g}\bm K$, either do
not overlap at all or overlap in a small way (Fig.~2) and, also, a
scattering of $\bm K$--pair from any such a domain into an
equivalent one corresponds to rather large change in the total
momentum of the pair, one can, in the first approximation, neglect
any inter-domain scattering. Then, with regard for Eqs.~(1), (2),
the equivalent Hamiltonian of the relative motion of $\bm K$--pair
may be presented in the form \cite{2}
\begin{equation}
\label{6}{\hat H}_K = -{\frac {\hbar^2}{2m}} {\left ({\nu }{\frac
{\partial{}^2 }{\partial {x_1^2} }}-{\frac {\partial{}^2}{\partial
{x_2^2}}}\right)} + U_K^{\ast }(r),
\end{equation}
where $r=\sqrt {x_1^2+x_2^2}$, $U_K^{\ast }(r)$ is an effective
potential energy of the particles composing $\bm K$--pair. This
energy, which will be defined and discussed in detail in Sec.~V,
depends on the domain $\Xi_K$ in which scattering due to
interaction is permitted. When the area $\Xi_K$ is large enough
one can suppose that \cite{2} $U_K^{\ast }(r)\sim \Xi_K$.

Generally speaking, all of the eigen-functions of the operator
Eq.~(6) belong to continuous spectrum. Therefore, it is quite
natural to represent such a function in the form of a sum of an
incident wave with the momentum  $\bm q$ and scattered (expanding)
wave,
\begin{equation}
\label{7}{\varphi_K}({\bm r}) \Rightarrow {\varphi_{Kq}}({\bm r})
= e^{i{\bm qr}} + {\chi_{Kq}}({\bm r}).
\end{equation}
Fourier transform of the scattered wave,
\begin{equation}
\label{8}{\tilde \chi_{Kq}}({\bm k}) = \int {\chi_{Kq}}(\bm
r){e^{-i{\bm kr}}}{d^2r},
\end{equation}
is a solution of the integral equation \cite{58}
\begin{eqnarray}
\label{9}&&[\omega -\omega ({\bm k})]{\tilde \chi_{Kq}}({\bm k})= \nonumber \\
&&=u({\bm k} - {\bm q}) + \int u({\bm k}-{\bm k^\prime }) {\tilde
\chi_{Kq}}({\bm k^\prime}){\frac {{d^2}{k^\prime}} {(2\pi)^2}}.
 \end{eqnarray}
Here, $\omega ({\bm k}) = {\nu k_1^2} - k_2^2$, ${\hbar^2 {\omega
}}/2m$  is an energy of the incident wave, $u({\bm k}) = 2m{\tilde
U_K^{\ast }}({\bm k})/{\hbar}^2$, ${\tilde U_K^{\ast }}({\bm k})$
is the Fourier transform of the effective interaction energy
introduced in Eq.~(6). One has to integrate in Eq.~(6) over the
domain $\Xi_K$ which is the domain of definition of momenta $\bf
k$ and $\bf k^\prime$. As far as this domain is small in
comparison with 2D Brillouin zone one can approximately take
$u({\bm k}-{\bm k^\prime}) \approx u(0)\equiv u_0$. This
approximation leads directly to the solution of the Eq.~(9) in the
form \cite{2},
\begin{equation}
\label{10}{\tilde \chi_{Kq}}({\bm k}) = -{\frac {u_0}
{1+u_0B_K(\omega )}}{\frac {1}{\omega({\bm k})-\omega -i0 \cdot
{\text{sgn}}\omega}},
\end{equation}
where signum function provides a necessary condition in order that
Eq.~(10) were an expanding wave. The function ${B_K}(\omega )$ is
defined as
\begin{eqnarray}
\label{11}  B_K(\omega)= && \int_{(\Xi_K)}{\frac 1{\omega({\bm
k})-\omega
-i0 \cdot {\text{sgn}\omega}}}{\frac {d^2k}{(2\pi)^2}}\nonumber\\
&& \qquad  \equiv B_{K1}(\omega)+iB_{K2}(\omega).
\end{eqnarray}
At real argument, the functions $B_{K1}(\omega)$ and
$B_{K2}(\omega)$ are written in the form
\begin{eqnarray}
\label{12}B_{K1}(\omega)=\int_{(\Xi_K)}{\frac 1{\omega ({\bm
k})-\omega}} {\frac {d^2k}{(2\pi)^2}},\nonumber\\
B_{K2}(\omega)={\pi }\cdot {\text{sgn} \omega}
\int_{(\Xi_K)}\delta (\omega ({\bm k})-\omega) {\frac
{d^2k}{(2\pi)^2}},
\end{eqnarray}
where the integral defining $B_{K1}(\omega)$ has meaning of
Cauchy principal value.

A denominator of the scattering amplitude,
\begin{equation}
\label{13}f_K(\omega)={\frac {u_0}{1+u_0B_K(\omega)}},
\end{equation}
generally speaking, is not equal to zero at any real value of the
argument $\omega $. The case when the function $B_{K2}(\omega)$
is equal to zero identically inside some interval of $\omega$ may
be considered as an exception. In such a case, scattering
amplitude poles, which are the solutions of the equation,
\begin{equation}
\label{14}1+u_0B_{K1}(\omega)=0,
\end{equation}
correspond to bounded states.

When some complex value, $\omega =\omega_K^{(0)} -i\Gamma_K$, is a
solution of Eq.~(14) and, in addition, $B_{K2}(\omega_K^{(0)})\neq
0 $, $\omega_K^{(0)}$ makes sense of the energy of
quasi-stationary state (QSS) provided that
$0<\Gamma_K<<\omega_K^{(0)}$. At $|\omega -\omega_K^{(0)}|<<
\omega_K^{(0)}$, the function $B_{K1}(\omega)$ may be
approximately written as
\begin{equation}
\label{15}B_{K1}(\omega) \approx
B_{K1}(\omega_K^{(0)})+B_{K1}^{\prime}(\omega_K^{(0)})\cdot
(\omega -\omega_K^{(0)}) .
\end{equation}
Here, the prime denotes differentiation with respect to $\omega$.
The scattering amplitude is represented in the form
\begin{equation}
\label{16}f_K(\omega)\approx {\frac
1{B_{K1}^{\prime}(\omega_K^{(0)})}}\cdot {\frac 1{\omega
-\omega_K^{(0)} +i\Gamma_K}} ,
\end{equation}
where QSS decay is written as
\begin{equation}
\label{17}\Gamma_K \approx B_{K2}(\omega_K^{(0)})/
B_{K1}^{\prime}(\omega_K^{(0)}) .
\end{equation}

In the case of tetragonal crystal (2D symmetry group $C_{4m}$),
one can separate all of the equivalent vectors  ${\hat g} \bm K$
into two subsets. One of them, which contains the vector $\bm K$
itself, also contains all of the vectors ${\hat g} \bm K$ related
to each other by reflections with respect to the coordinate axes,
$k_x$ and $k_y$. The another subset is generated in a similar way
by the vector resulting from the reflection of $\bm K$  with
respect to a diagonal of the square Brillouin zone. The
coefficients $c_{{\hat g}K}^{(A_{1g})}$ corresponding to the
trivial irreducible representation $A_{1g}$ are equal to each
other. In the case of the irreducible representation $B_{1g}$, the
coefficients $c_{{\hat g}K}^{(B_{1g})}$ have one and the same
absolute value and differ in sign for the two subsets of the full
set ${\hat g} \bm K$. Taking into account the explicit form,
Eq.~(10), of the functions ${\tilde \chi}_{{\hat g}Kq}({\bm k})$
one may easily conclude that, in the case of an appropriate choice
of coordinate axis directions ${\omega}({\bm k})=\nu k_x^2-k_y^2$,
for any ${\hat g} \bm K$ belonging to the first subset of the full
set of the vectors, ${\hat g} \bm K$, whereas ${\omega}({\bm
k})=\nu k_y^2-k_x^2$ when ${\hat g} \bm K$ belong to the second
one. Thus, the wave function Eq.~(5) corresponding to the
irreducible representation $A_{1g}$ has the form
\begin{equation}\label{18}
\Psi_{Kq}^{(A_{1g})} \sim {\frac {2\omega}{(\omega + k_x^2)\cdot
(\omega +k_y^2)}} ,
\end{equation}
provided that $\nu <<1 $. Full symmetry of this function with
respect to the crystal group enables one to relate it to
$s$--type orbital symmetry. Under the same condition, $\nu <<1 $,
the wave function corresponding to the irreducible representation
$B_{1g} $ may be written as
\begin{equation}
\label{19} \Psi_{Kq}^{(B_{1g})} \sim {\frac {k_x^2-k_y^2}{(\omega
+ k_x^2)\cdot (\omega +k_y^2)}}.
\end{equation}
This function may be conditionally related to $d$--type orbital
symmetry.

In the coordinate representation, the wave function of the
relative motion of $\bm K$--pair corresponding to an expanding
wave, at $\omega >0$, has the form
\begin{eqnarray}
\label{20}\chi_{Kq}({\bm r})={\frac {f_K(\omega)}{4\sqrt\nu}}\cdot
\left\{\aligned H_0^{(2)}({\sqrt
\omega}{\sqrt {{x^2/\nu}-y^2}}),\\
{\frac {2i}{\pi}}K_0({\sqrt{ \omega}}{\sqrt
{y^2-{x^2/\nu}}}).\endaligned \right .
\end{eqnarray}
Here, $H_0^{(2)}(z)$ and $K_0(z)$ are Hankel and modified Bessel
functions, respectively. The upper (low) row corresponds to
$0<|y|<|x|/\sqrt{\nu}$ ($|x|/\sqrt{\nu}<|y|<\infty$). Thus, the
plane $x, y$ is separated into four sectors. Two sectors, namely,
$0<|y|<|x|/\sqrt \nu$, correspond to expanding waves whereas in
the another two, $|x|/\sqrt \nu <|y|<\infty$, the wave function,
Eq.~(20), decays with the distance from the center. The straight
lines, $y=\pm x/\sqrt \nu$, separating these sectors are caustics
on which the function, Eq.~(20), has a logarithmic singularity. At
$\omega <0$, we have
\begin{eqnarray}
\label{21}\chi_{Kq}({\bm r})=-{\frac
{f_K(\omega)}{4\sqrt\nu}}\cdot\left\{\aligned H_0^{(2)}({\sqrt {-
\omega}}{\sqrt {y^2-{x^2/\nu}}}),\\
{\frac {2i}{\pi}}K_0({\sqrt {-\omega}}{\sqrt
{y^2-{x^2/\nu}}}). \endaligned \right .
\end{eqnarray}
Here, the upper (low) row corresponds to $0<{|x|/\sqrt\nu}<|y|$
($|y|<{|x|/\sqrt\nu}<\infty$). A character of coordinate
dependence of this function, that is a disposition of the sectors
corresponding to expanding and damping waves, is clear from the
definition, Eq.~(21). It should be noted that Eqs.~(20), (21) are
related to the case when the energy of the relative motion of $\bf
K$--pair is measured from the center--of--mass energy. One can
easily see that a character spatial scale for the pairs considered
here (which plays role of a coherence length) is determined by the
value of the order of $|\omega|^{-1/2}$ and may be estimated as a
few interatomic distances \cite{2}.


\section{ Quasi-stationary states}

One can calculate the function Eq.~(12) which defines the
scattering amplitude. For the sake of simplicity, we suppose that
the domain $\Xi_K$  being the domain of integration in Eq.~(12) is
a long and narrow rectangular strip which is roughly similar to
real domain  $\Xi_K$ in the case of antinodal direction. We denote
a length and a width of the strip as $\Delta k_1$  and $\Delta
k_2$, respectively. Coordinate $k_1$--axis is directed along one
of the principal directions of 2D reciprocal effective mass tensor
which corresponds to positive effective mass, $m_1=m/\nu$. Another
axis, $k_2$, is directed along the principal direction
corresponding to negative effective mass, $m_2=-m$. Using such an
approximation, one can represent the function $B_{K2}(\omega)$ at
$\omega>0$ in the explicit form,
\begin{eqnarray}
\label{22}B_{K2}(\omega)={\frac 1{2\pi \sqrt\nu}}\cdot
\left\{\aligned \ln \left |{\frac {{\sqrt
{\omega_{-1}}}+{\sqrt{\omega_{-1}+\omega}}}{\sqrt\omega}}\right |,\\
\ln \left |{\frac {{\sqrt
{\omega_{+1}}}+{\sqrt{\omega_{+1}-\omega}}}{\sqrt\omega}}\right |,
\endaligned \right .
\end{eqnarray}
where $\omega_{-1}=(\Delta k_2/2)^2$, $\omega_{+1}={\nu}(\Delta
k_1/2)^2$. The upper (low) row corresponds to
$0\leq\omega_{-1}\leq\omega_{+1}-\omega $
($0\leq\omega_{+1}-\omega\leq\omega_{-1}$). The function Eq.~(22)
is defined inside the interval $0\leq\omega\leq\omega_{+1}$ and
depends on both $\omega_{-1}$ and $\omega_{+1}$. Taking into
account that $\nu<<1$ and $\Delta k_1>>\Delta k_2$ one may assume,
for the sake of simplicity, that
$\omega_{-1}=\omega_{+1}\equiv\omega_1$. This assumption using
later on is sufficient to study the main features of the
scattering amplitude. Thus, one can see that the upper solution in
Eq.~(22) has not a domain of definition whereas the lower one is
defined inside the whole of the interval
$0\leq\omega\leq\omega_1$. Similar consideration provided that
$\omega<0$ leads to the explicit expression for $B_{K2}(\omega)$
at any $\omega$,
\begin{equation}\label{23}
 B_{K2}(\omega)=\frac {{\text{sgn}}\omega}{2\pi \sqrt
{\nu}}\cdot \ln{\left |\frac {\sqrt {\omega_1-|\omega|}+{\sqrt
{\omega_1}}}{\sqrt {|\omega|}}\right |}.
\end{equation}
Here, $0\leq |\omega|\leq \omega_1$. It should be noted that, at
$|\omega|>\omega_1$, we have $B_{K2}(\omega)=0$.

The function $B_{K2}(\omega)$ is connected with the density of
states of the relative motion of $\bm K$--pair inside the domain
$\Xi_K$,
\begin{equation}
\label{24}B_{K2}(\varepsilon)=\pi \varepsilon_0 \cdot
g_K(\varepsilon) \cdot \text{sgn} {\varepsilon} ,
\end{equation}
where $\varepsilon_0=({\hbar}^2\omega_1/m)$ is an energy width of
the domain $\Xi_K$, $a^2=S/N$, $N$ is a number of unit cells in
conducting plane, $\varepsilon=({\hbar}^2\omega/m)$. Average
(inside the domain $\Xi_K$) density of states per unit cell is
defined as
\begin{equation}
\label{25}g_K={\frac 1{\varepsilon_0}}
\int_{-\varepsilon_0/2}^{\varepsilon_0/2} g_K(\varepsilon),
d\varepsilon .
\end{equation}
One can rewrite Eq.~(25) in the form
\begin{equation}
\label{26}g_K={\frac 1{\varepsilon_0}}\cdot {\frac {\Xi_K a^2}
{(2\pi)^2}} = {\frac {ma^2}{{\pi}^2{\hbar}^2}}\cdot {\frac
1{\sqrt \nu}} .
\end{equation}
Due to the condition that $\nu<<1$, average density of pair
states inside the domain $\Xi_K$  may be considerably more in
comparison with total average (inside 2D Brillouin zone) density
of states which is equal to $ma^2/{\pi}^2{\hbar}^2$. This is a
consequence of peculiar features of saddle point vicinity in HTSC
cuprates (known as extended saddle point) associated with
hyperbolic metrics and strong effective mass anisotropy.

The function defined by Eq.~(23) has a logarithmic singularity at
$|\omega|\rightarrow 0$,
\begin{equation}
\label{27}B_{K2}(\omega) \sim {\frac {\text{sgn}\omega}{4\pi\sqrt
\nu}}\ln{\left |{\frac {4\omega_1}{\omega}}\right |} ,
\end{equation}
that is the well-known, typical of any 2D system, logarithmic van
Hove singularity in density of states due to saddle points. Near
the edges of the energy band, $-\omega_1\leq\omega\leq\omega_1$,
in which the function $B_{K2}(\omega)$ is defined, this function
behaves as follows:
\begin{equation}
\label{28}B_{K2}(\omega) \sim {\frac {\text{sgn}
\omega}{2\pi\sqrt \nu}}{\sqrt{1-|\omega|/\omega_1}} , \quad\
|\omega_1-|\omega||<<\omega_1 .
\end{equation}
The function $B_{K2}(\omega)$, Eq.~(23), is plotted in Fig.~5.

Now, let us consider the function $B_{K1}(\omega)$ defined in
Eq.~(12). Taking into account Eq.~(24) we have
\begin{eqnarray}
\label{29}\aligned B_{K1}(\omega)=\int {\frac 1{\nu k_1^2 - k_2^2
-\omega}} {\frac {d^2k}{(2\pi)^2}}\\
= {\frac 1{\pi}}\int_{-\omega_1}^{\omega_1} {\frac
{B_{K2}(\omega^\prime) \cdot {\text{sgn}
\omega^\prime}}{\omega^\prime -\omega}} d\omega^\prime.
\endaligned
\end{eqnarray}
First of all, let us estimate $B_{K1}(\omega)$ using the average
value, Eq.~(26), of pair density of states. We obtain
\begin{equation}
\label{30}B_{K1}(\omega)\approx {\frac 1{2{\pi}^2\sqrt
\nu}}\int_{-\omega_1}^{\omega_1}{\frac {d\omega}{\omega^\prime
-\omega}} = {\frac 1{2{\pi}^2\sqrt \nu}}\ln{\left |{\frac
{\omega_1-\omega}{\omega_1+\omega}} \right |}.
\end{equation}
The function Eq.~(30) is presented in Fig.~6 (dashed line). It
should be noted that just the function of the form Eq.~(30) was
used \cite{47} to analyze the ARPES experiment. More rigorous
treatment of such self-energy structure was done in \cite{47'}
from the analysis of energy and momentum distribution curves. The
explicit expression of $B_{K1}(\omega)$ ,
\begin{equation}
\label{31}B_{K1}(\omega)={\frac {\omega}{{\pi}^2{\sqrt \nu}}}
\int_0^\infty \ln{\left |{\frac {{\sqrt
{\omega_1-\omega^\prime}}+{\sqrt \omega_1}}{\sqrt
{\omega^\prime}}}\right |}  {\frac
{d\omega^\prime}{{\omega^\prime}^2 - \omega^2}} ,
\end{equation}
in the form of a combination of elementary or special functions
is unknown. It is obvious that, at $|\omega|\rightarrow \infty$
\begin{equation}
\label{32}B_{K1}(\omega) \sim - {\frac {\Xi_K}{(2\pi)^2}}\cdot
{\frac 1{\omega}} , \qquad\ |\omega|\rightarrow \infty .
\end{equation}
At $\omega \rightarrow \pm0$, we have from the definition,
Eq.~(29),
\begin{equation}
\label{33}B_{K1}(\pm 0) = \pm {\frac 1{2\pi^2{\sqrt \nu}}}
\int_0^1 \ln{\left |{\frac {1-x}{1+x}}\right |} {\frac {dx}{x}} =
\pm {\frac 1{8 \sqrt \nu}}
\end{equation}
Thus, a logarithmic singularity of density of states appears in
$B_{K1}(\omega)$ as a finite discontinuity at $\omega \rightarrow
\pm0$. At $\omega = \pm \omega_1$, the function $B_{K1}(\omega)$
has the finite values,
\begin{eqnarray}
\label{34} B_{K1}(\pm \omega_1)= &&  \mp{\frac 1{\pi^2{\sqrt
\nu}}} \int_0^1 \ln{\left |{\frac {\sqrt
{1-x}+1}{\sqrt x}}\right |} {\frac {dx}{1-x^2}} \nonumber\\
&& \approx \mp
{\frac {0.164}{\sqrt \nu}},
\end{eqnarray}
with $|B_{K1}(\pm \omega_1)|>|B_{K1}(\pm 0)|$. The function
$B_{K1}(\omega)$, Eq.~(31), is plotted in Fig.~6 (solid line).

The obtained function, $B_{K1}(\omega)$, allows us to analyze
qualitatively the solutions of the Eq.~(14) which determines the
poles of the scattering amplitude. In the case of repulsion
between particles composing $\bf K$--pair ($u_0>0$) the solutions
corresponding to positive energy ($\omega >0$) exist provided
that $-u_0 B_{K1}(\omega_1)<1$. One of the solutions,
$\omega^+_{QSS}$, corresponding to greater energy, exists at as
much as desired value of the coupling constant, $u_0$. The second
solution, $\omega^+_{SC}$, exists in a bounded interval of
coupling constant values: $|B_{K1}^{-1}(+0)| < u_0 <
|B_{K1}^{-1}(\omega_1)|$. The first solution takes place when
$\omega_1 < \omega^+_{QSS} < \infty$ and the second one exists in
an energy interval bounded both above and below, $0 <
\omega^+_{SC} < \omega_1$. At $\omega > 0$, as it is shown in
Fig.~5, $B_{K2}(\omega)>0$, therefore, the sign of a decay which
corresponds to both poles of the scattering amplitude is
determined by the behavior of the function $B_{K1}(\omega)$. In
the interval $\omega_1 < \omega < \infty$, this function
increases with $\omega$, therefore, positive decay,
$\Gamma^+_{QSS} > 0$, corresponds to the pole $\omega^+_{QSS}$.
Hence, this pole may be really associated with a quasi-stationary
state. But in fact, as it follows from the definition, Eq.~(12),
and Fig.5, $\Gamma^+_{QSS} = + 0$, therefore, the approximation
used here leads to the pole $\omega^+_{QSS}$  being a real
stationary state. Indeed, one can see that, due to hyperbolic
metrics of momentum space, at $\omega
> \omega_1$, any decomposition of $\bm K$--pair becomes
impossible because of the restrictions connected with momentum
and energy conservation.

At $0 < \omega <\omega_1$ the function $B_{K1}(\omega)$, on the
contrary, decreases with $\omega$, therefore, finite and negative
decay, $\Gamma^+_{SC}<0$, corresponds to the pole. This fact may
be considered as an evidence of an instability with respect to a
rise of $\bm K$--pairs, and imaginary part of the pole,
$\Gamma^+_{SC}$, may be directly connected with SC gap in
one-particle excitation spectrum. However, the presence of
positive real part of the pole, $\omega^+_{SC} > 0$, indicate that
a rise of the SC state becomes possible only if an energy increase
connected with the finite value of $\omega^+_{SC}$ were
compensated by sufficient energy decrease produced by
corresponding rearrangement of the electron system which does not
bear a direct relation to $\bf K$--pair formation. As an example
of such a rearrangement in HTSC cuprates, one may consider above
mentioned rise of spatially inhomogeneous spin and charge
structure because of partial restoration of AF order. Thus,
because of the positive sign of the real part of the scattering
amplitude, the SC pole without any renormalization of the ground
state may be considered as corresponding to a metastable state.
The QSS state has to be related not to a minimum but a maximum of
total energy.

In the case when attraction between particles composing $\bm
K$--pair dominates ($u_0<0$), at $0 < |u_0| < B_{K1}(-\omega)$, as
it is seen from Fig.6, there is a solution, $\omega^-_{SC}$, of
Eq.~(14) which exists in infinite energy region, $-\infty <
\omega^-_{SC} < -\omega_1$, and, in the approximation used here,
has an infinitesimal decay, $\Gamma^-_{SC} = -0$. On the contrary,
another pole, $\omega^-_{QSS}$, existing inside the energy
interval, $-\omega_1 < \omega^-_{QSS} < 0$, bounded both above and
below, corresponds to real QSS with finite and rather large decay
as one can see from Fig.~ 5. Therefore, in spite of the fact that,
due to assumption that $\omega_{-1} = \omega_{+1}$, there is
obvious symmetry of the function $B_{K1}(\omega)$ with respect to
a change of the sign of the argument, namely, $B_{K1}(-\omega) = -
B_{K1}(\omega)$, there is essential asymmetry in a character of
solutions of Eq.~(14) with respect to the sign of the coupling
constant.


\section{ Interaction between particles composing $\bm K$-- pair}

The point of view \cite{59,60,61} that there are incoherent
electron or hole pairs in the pseudogap regime leads to definite
conclusion concerning the sign of the interaction energy which
governs the pairing in HTSC cuprates. Namely, one may propose
that the only essential interaction between electrons is screened
Coulomb repulsion,
\begin{equation}
\label{35}U(r)={\frac {e^2}{r}}\cdot \exp{\left (-{\frac
r{r_0}}\right )} ,
\end{equation}
where $r_0 = [4\pi e^2 N(E_F)]^{-1/2}$ is a screening length,
$N(E_F)$ is one-particle density of states on the Fermi level,
$E_F$. The potential (5.1) may be expanded into Fourier integral
or discrete Fourier series inside 2D Brillouin zone, $Z_B$:
\begin{equation}
\label{36}U(r)=\int_{Z_B} {\tilde U}({\bm k}) e^{i{\bm {kr}}}
{\frac {d^2k}{(2\pi)^2}}={\frac 1{S}}\sum_{k\in Z_B} {\tilde
U}({\bm k}) e^{i{\bm {kr}}}
\end{equation}
Here, ${\tilde U}(\bm k)$ is the Fourier transform of the
potential Eq.~(35), $S$ is a normalizing area. A momentum $\bm k$
of the relative motion of a pair with total momentum $\bm K$
belongs to some domain $\Xi_K$ inside the Brillouin zone.
Therefore, to describe the effective interaction, $U_K^{\ast}(r)$,
of the particles composing a pair one can use the expression
Eq.~(36) in which, however, the only essential momenta of the
relative motion are those belonging to corresponding domain
$\Xi_K$:
\begin{equation}
\label{37}U_K^{\ast}(r)=\int_{\Xi_K} {\tilde U}({\bm k}) e^{i{\bm
{kr}}}{\frac {d^2k}{(2\pi)^2}}={\frac 1{S}}\sum_{k\in \Xi_K}
{\tilde U}({\bm k}) e^{i{\bm {kr}}}.
\end{equation}
If the domain $\Xi_K$ includes many quantum states and, on the
other hand, this domain is small enough in comparison with the
Brillouin zone area,
\begin{equation}
\label{38}{\frac {(2\pi)^2}{S}}<<\Xi_K<< {\frac {(2\pi)^2}{a^2}} ,
\end{equation}
one can easily estimate the expression Eq.~(37) using the theorem
about the mean value. Let us suppose that ${\tilde U}({\bm k})
\approx {\tilde U}(0)$ inside the whole of the domain $\Xi_K$.
Then, at $r \neq 0$, oscillating terms in the sum, Eq.~(37),
suppress each other essentially. Thus, we have approximately,
\begin{equation}\label{39}
U_K^{\ast}(r)\approx U_K^{\ast}(0)\cdot {\delta_{{\bm r}0}} ,
\end{equation}
where
\begin{equation}
\label{40}U_K^{\ast}(0)={\frac 1{S}} \sum_{k\in \Xi_K} {\tilde
U}({\bm k}) = {\tilde U}(0){\frac {\Xi_K}{(2\pi)^2}} .
\end{equation}
Taking account that
\begin{equation}
\label{41}\delta_{{\bm r}0}=\sum_{k\in Z_B} e^{i {\bm {kr}}}
\Rightarrow a^2 \delta({\bm r}) ,
\end{equation}
where $N$ is a number of unit cells corresponding to the area $S$,
$a$ is an interatomic distance, one can conclude that the
effective interaction between the particles composing a pair with
total momentum $\bm K$  may be approximately considered as a
contact interaction of the form
\begin{equation}
\label{42}U_K^{\ast}(0)\approx {\tilde U}(0) {\frac {\Xi_Ka^2}
{(2\pi)^2}} \delta ({\bm r}).
\end{equation}
The Fourier transform of the effective interaction Eq.~(42) has
the form
\begin{equation}
\label{43}{\tilde U}_K^{\ast}(k)={\tilde U}(0){\frac
{\Xi_Ka^2}{(2\pi)^2}}={\frac {e^2r_0\Xi_Ka^2}{2\pi}}.
\end{equation}
Here we take into account the explicit form of the Fourier
transform of the screened Coulomb potential Eq.~(35), ${\tilde
U}(k) =(2\pi e^2 r_0)/(1+k^2r^2_0)$. Now, the coupling constant
may be written in the form
\begin{equation}\label{44}
u_0={\frac {r_0}{\pi a^{\ast}}}\Xi_Ka^2
\end{equation}
where $a^{\ast} = {\hbar}^2/me^2$ is an effective Bohr radius. An
increase of carrier concentration due to doping leads to a
decrease of the screening length and, as a result, to a decrease
of the coupling constant. In Fig.~7, we represent a qualitative
comparison of typical of HTSC cuprates phase diagram (Fig.1) and
Fig.~6 which we consider here as a dependence of $\bf K$--pair
energy $\omega$  on, increasing with doping, inverse value of the
coupling constant, $u_0^{-1}$. One can see from Fig.~7 that a
crossover line, $T^{\ast}(x)$, separating normal and pseudogap
states in phase diagram is in accordance with the line which
determines the energy of QSS with positive infinitesimal decay,
${\omega}^+_{QSS}$. Besides, there is an obvious accordance
between SC phase region bounded by the line $T_C(x)$ in phase
diagram and the line which determines the solution leading to SC
instability, ${\omega}^+_{SC}$. Indeed, as it is clearly seen from
Fig.~7, both functions of doping, $T_C(x)$ and
${\omega}^+_{SC}(x)$, have finite domains of definition bounded
above and below.


\section{ Superconducting pairing}

The presence of negative-decay poles in the scattering amplitude
corresponding to the relative motion of electron or hole pair with
large (of about $2k_F$) total momentum bears evidence to a
possibility of SC pairing both at attraction and repulsion between
the particles composing the pair. A consequence of a rise of
spatially inhomogeneous electron structure such as stripe
structure is that a number of real $\bm K$--pairs belonging to the
domain $\Xi_K$  must leave this domain and form new real pairs
with a momentum  ${\bm K}^\prime$ ( ${\bm K}^\prime$--pairs) in a
domain $\Xi_{K^\prime}$ outside of the FC. The states inside
$\Xi_K$ and $\Xi_{K^\prime}$ having become vacant and remaining
filled are separated from each other by a line which is, by our
definition, the PFC. The area $\Xi_K^{(+)}$ corresponding to
vacant states inside $\Xi_K$ is, generally speaking, not equal to
the area $\Xi^{(-)}_{K^\prime}$ of the filled part of the domain
$\Xi_{K^\prime}$, if one takes into account the fact that, in a
general case, the areas of AF and M parts of a stripe not equal to
each other. The values of each of the areas, $\Xi_K^{(+)}$ and
$\Xi^{(-)}_{K^\prime}$ are dependent on the AF energy which
determines the position of the chemical potential $2\mu$ of pairs
with respect to the edges of the energy bands corresponding to the
domains $\Xi_K$ and $\Xi_{K^\prime}$. These energy bands and
relevant densities of states, $g_K(\varepsilon)$ and
$g_{K^\prime}(\varepsilon)$, are represented schematically in
Fig.~3. One can see that one part of the PFC, which is situated in
the domain $\Xi_K$ (the boundary between $\Xi_K^{(-)}$ and
$\Xi_K^{(+)}$), may be related to AF part of a stripe whereas
another part, separating $\Xi_{K^\prime}^{(-)}$ and
$\Xi_{K^\prime}^{(+)}$, belongs to M part of a stripe. An
excitation of carriers (a rise of holes above and electrons below
the chemical potential level) leads to a possibility of their
pairing. Formally, one can consider (1) a scattering of pairs in
AF part of a stripe (in the domain $\Xi_K$), (2) a scattering of
pairs in M part of a stripe (in the domain $\Xi_{K^\prime}$), and
also, (3) a scattering which includes transfers of pairs between
AF part of a stripe (the domain $\Xi_K$) and M part of a stripe
(the domain $\Xi_{K^\prime}$) as it is in the case discussed below
in Sec.~X. In such a case, (3), pairs are spatially separated and
the interaction leading to their scattering is reduced. As it is
already mentioned, the condition $|{\bm K^\prime} - {\bm K}| >>
\delta k_c$ allows us, in the first approximation, to consider
pairing in the domains $\Xi_K$ and $\Xi_{K^\prime}$ independently
of each other thus restricting ourselves to one of the cases (1)
or (2). It should be noted that the equation determining the SC
order parameter has a set of solutions and, among them, there are
solutions which, in the cases (1) and (2), may turn out to be
trivial (corresponding order parameter becomes equal to zero).
However, in such a case, a nontrivial solution may arise just due
to the third, (3), possibility (for example, in the case of weak
ferromagnetic ordering, \cite{7}). In this Section, we restrict
ourselves to the case (1) and consider SC pairing near the part of
the PFC belonging to AF part of a stripe. Thus, we suppose that
rather thin M parts of stripes do not affect the superconductivity
essentially due to the proximity effect.

In general case, considering a pairing of carriers with momenta
$\bm K$ or $\bm K^\prime$ along an antinodal direction, it is
necessary to take into account all excited states arising due to
transfers of carriers across the PFC, namely,
$\Xi_K^{(-)}\leftrightarrow \Xi_K^{(+)}$, $\Xi_{K^\prime}^{(-)}
\leftrightarrow \Xi_{K^\prime}^{(+)}$ (intradomain excitations)
and $\Xi_{K^\prime}^{(-)}\leftrightarrow \Xi_K^{(+)}$,
$\Xi_K^{(-)}\leftrightarrow \Xi_{K^\prime} ^{(+)}$  (interdomain
excitations). As a result, we have a pair with large total
momentum along one of the antinodal directions. As it is mentioned
above (Sec.~III), in the case of the antinodal directions, there
exists a quadruple of equivalent pairs with total momenta ${\hat
g} \bm K$ where, due to symmetry, $\sum_{\hat g} {{\hat g} \bm K}
= 0$. The interaction leading to a scattering such equivalent
pairs turns out to be more weak as compared to the interaction
which results in a rise of a bound state of a pair with given $\bf
K$. This may be, mainly, due to an essential increase of the
scattering momentum in spite of the fact that the scattering
region in momentum space increases too. Scattering of pairs with
equivalent total momenta leads to a state of the form Eq.~(5)
which, due to the condition that $\sum_{\hat{g}} {{\hat{g}} \bm K}
= 0$, corresponds to a current-less state. In this sense, a pair
state as a quadruple of pairs with equivalent large total momenta
is similar to a conventional Cooper pair, however, it is clear
that the internal structure of pair states here discussed differs
essentially from rather simple structure of Cooper pair.

It is obvious that stripe structure periodicity in a conducting
plane and the difference $|{\bm K^\prime}-{\bm K}|$ have to be
correlated. For simple 1D stripe structure, ${\bm K}$ (${\bm
K^\prime}$)--pair state arises due to a mixing of only two ${\bm K
}$--states corresponding to either [100] ($k_1$--axis) or [010]
($k_2$--axis) direction. Thus, one may expect a rise of an array
of alternating $CuO_2$ planes with 1D stripes which are
perpendicular to each other in the neighboring planes. The pair
states formed by the quadruples of ${\bm K}$--pairs correspond to
more complicated periodic 2D stripe structure. Real (nonperiodic)
AF short-range-order fluctuations may be described as linear
combination of ${\bm K}$ (${\bm K^\prime}$)--pair as well.

The experimental data \cite{29,30} bear evidence to singlet
pairing in HTSC cuprates. Therefore, we write the equivalent
Hamiltonian corresponding to relative motion of $\bm K$--pairs in
the form
\begin{eqnarray}\label{45}
\hat H_K = && \sum_k {\left [ (\varepsilon_{{\bm k}_+}-{\mu})
{\hat{a}^{\dag}_{k_+\uparrow}}{\hat{a}^{}_{k_+\uparrow}}
 + (\varepsilon_{{\bm k}_-}-\mu) {\hat{a}^{\dag}_{k_-\downarrow}}
{\hat{a}^{}_{k_-\downarrow}}\right ]}\nonumber\\
&& +{\frac 1{S}} \sum_{k,k^\prime} {\tilde U}_K^{\ast}({\bm k}-
{\bm k^\prime}) {\hat{a}^{\dag}_{k_+\uparrow}}
{\hat{a}^{\dag}_{k_-\downarrow}}
{\hat{a}^{}_{{{k^\prime}_-}\downarrow}}
{\hat{a}^{}_{{{k^\prime}_+}\uparrow}},
\end{eqnarray}
where $\varepsilon_{{\bm k}_\pm} \equiv \varepsilon(\bm k_\pm)$ is
hole dispersion, ${\tilde U}_K^{\ast}({\bm k} - {\bm k^\prime})$
is Fourier transform of the effective interaction energy,
Eq.~(37), ${\bm k^\prime}_{\pm} = {\bm K}/2 \pm {\bm k^\prime}$,
${\hat a}_{k_{\pm}\sigma}^{\dag}$ (${\hat a}_{k_{\pm}\sigma}^{}$)
creates (annihilates) a hole with a momentum ${\bm k}_{\pm}$ and
spin quantum number $\sigma$. The symbol $\uparrow$ ($\downarrow$)
is referred to $\sigma = 1/2$ ($\sigma = -1/2$), $\mu$ is a hole
chemical potential. The summation in Eq.~(45) is taken over all
range of values of momenta of the relative motion of $\bm
K$--pair. Note that the summation in the Hamiltonian Eq.~(45)
should be taken over only two (instead of three in general case)
variables, $\bm k$ and ${\bm k}^\prime$, just as in the case of
BCS Hamiltonian, thus taking into account the only interaction
between particles composing $\bm K$--pairs. Also, it should be
noted that the summation in Eq.~(45) should be restricted by the
corresponding domain $\Xi_K$ as it is discussed in Section~V.

As usual, to diagonalize the Hamiltonian Eq.~(45) approximately
one can introduce creation and annihilation operators of new
quasiparticles using the well-known Bogoliubov-Valatin
transformation \cite{62,63}:
\begin{eqnarray}
\label{46}
{\hat{a}_{k_+\uparrow}^{}}=u_{Kk}
{\hat{b}_{k,{+1}}^{}}+ v_{Kk} {\hat{b}^{\dag}_{k,{-1}}} ,\nonumber\\
{\hat{a}^{\dag}_{k_+\uparrow}}=u_{Kk}
{\hat{b}^{\dag}_{k,{+1}}}+ v_{Kk} {\hat{b}_{k,{-1}}^{}} ,\nonumber\\
{\hat{a}_{k_-\downarrow}^{}}=u_{Kk}
{\hat{b}_{k,{-1}}^{}}- v_{Kk} {\hat{b}^{\dag}_{k,{+1}}} ,\nonumber\\
{\hat{a}^{\dag}_{k_-\downarrow}}=u_{Kk} {\hat{b}^{\dag}_{k,{-1}}}-
v_{Kk} {\hat{b}_{k,{+1}}^{}}.
\end{eqnarray}
The Hamiltonian, up to the terms of the order of ${\hat b}^2$, is
written as
\begin{equation}
\label{47}{\hat H}_K^{} =E_{K0}^{} + {\hat H}_K^{(0)} + {\hat
  H}_K^{(1)}.
\end{equation}
The ground state energy has the form
\begin{equation}
\label{48}E_{K0}^{}=2\sum_k \xi_{Kk}^{}v^2_{Kk} + \sum_k
\Delta_{Kk}^{} u_{Kk}^{}v_{Kk}^{},
\end{equation}
where, related to the value of the chemical potential, an energy
of the relative motion of $\bm K$--pair is defined as
\begin{equation}
\label{49}2\xi_{Kk}^{}=\varepsilon({\bm k}_+) +\varepsilon({\bm
k}_-) -2\mu .
\end{equation}
Diagonal, with respect to quasiparticle operators, part of the
Hamiltonian is given by
\begin{equation}
\label{50}\hat{H}_K^{(0)} = \sum_{k;\beta =\pm 1} \eta_{K\beta}^{}
({\bm k}) {\hat{b}^{\dag}_{k,{\beta}}}{\hat{b}_{k,{\beta}}^{}} ,
\end{equation}
where the energies corresponding to two branches ($\beta =\pm1$ )
of one-particle excitation spectrum are equal to each other,
\begin{equation}
\label{51}\eta_{K\beta}^{}({\bm k}) = \sqrt {\xi^2_{Kk} +
\Delta^2_{Kk}} .
\end{equation}
Nondiagonal, with respect to quasipartice operators, part of the
Hamiltonian has the form
\begin{eqnarray}
\label{52}\hat{H}_K^{(1)}= && \sum_k \left [2\xi_{Kk}^{}
u_{Kk}^{}v_{Kk}^{} -
(v^2_{Kk} - u^2_{Kk})\Delta_{Kk}^{} \right ] \nonumber\\
&& \times ({\hat{b}^{\dag}_{k,{+1}}}{\hat{b}^{\dag}_{k,{-1}}} +
{\hat{b}_{k,{-1}}^{}}{\hat{b}_{k,{+1}}}^{}) .
\end{eqnarray}
Here we define the order parameter as
\begin{equation}
\label{53}\Delta_{Kk}^{} = {\frac 1S} \sum_{k^\prime}{\tilde
U}_K^{\ast} ({\bm k}-{\bm k^\prime}) u_{Kk^\prime}^{}
v_{Kk^\prime}^{} \cdot f_{k^\prime}^{},
\end{equation}
where
$f_{k^\prime}^{}=(1-n_{{k^\prime},+1}^{}-n_{{k^\prime},-1}^{})$,
\begin{equation}\label{54}
n_{k\beta}^{} \equiv {\langle
{{\hat{b}^{\dag}_{k,\beta}}{\hat{b}_{k,\beta}^{}}} \rangle}
={\frac {1}{\exp(\eta_{K\beta}^{}(k)/T)+1}}
\end{equation}
are quasiparticle occupation numbers.

A choice of the amplitudes in Bogoliubov-Valatin transformation
Eq.~(46) is determined by the conditions that, in the
zero-temperature limit, the subdomain $\Xi_K^{(-)}$, in which
kinetic energy of the relative motion of $\bm K$--pair is
negative, $2\xi_{Kk} < 0$, must be filled and the nondiagonal
part, Eq.~(52), of the Hamiltonian vanishes. In addition, the
condition $u^2_{Kk} + v^2_{Kk} = 1$ preserving Fermi's commutation
relations for quasiparticle operators must be fulfilled. These
conditions yield
\begin{eqnarray}
\label{55} && v^2_{Kk}={\frac 12}\left (1- \frac
{\xi_{Kk}^{}}{\sqrt
{{\xi^2_{Kk}}+{\Delta^2_{Kk}}}} \right ), \nonumber\\
&&  u_{Kk}^{}v_{kk}^{} = -{\frac 12}{\frac {\Delta_{Kk}^{}}{\sqrt
{\xi^2_{Kk}+\Delta^2_{Kk}}}}.
\end{eqnarray}
Now, one can obtain the equation determining the order parameter,
\begin{equation}
\label{56}\Delta_{Kk}^{}=-{\frac 1{2S}}\sum_{k^\prime}{\frac
{{\tilde U}_K^{\ast}({\bm k}-{\bm
k^\prime})\Delta_{Kk^\prime}^{}}{\sqrt
{\xi^2_{Kk^\prime}+\Delta^2_{Kk^\prime}}}}\cdot f_{k^\prime}^{},
\end{equation}
where summation is taken over the whole of the domain $\Xi_K$. It
should be noted that kinetic energy of the relative motion of $\bm
K$--pair inside $\Xi_K^{(+)}$ is defined within the limits $0 \leq
2\xi_{Kk} \leq 2\varepsilon_{K+}$ where $\varepsilon_{K+}$ is an
energy width of the subdomain $\Xi_K^{(+)}$; inside $\Xi_K^{(-)}$
we have $-2\varepsilon_{K-} \leq 2\xi_{Kk} \leq 0$ where
$\varepsilon_{K-}$ is an energy width of the subdomain
$\Xi_K^{(-)}$ (Fig.~3). Inside the whole of the domain $\Xi_K$,
the excitation energy Eq.~(51) is positive by definition,
$\eta_{\pm 1}^{}(k) = \sqrt {\xi^2_{Kk} + \Delta^2_{Kk}} > 0$, and
hence, in the zero-temperature limit, the factor $f_k=1$ at any
$\bm k$ belonging to $\Xi_K$. Therefore, in the zero-temperature
limit, the equation Eq.~(56) transforms into the form
\begin{equation}
\label{57}\Delta_{Kk}=-{\frac 1{2S}}\sum_{k^\prime}{\frac {{\tilde
U}_K^{\ast}({\bm k}-{\bm k^\prime})\Delta_{Kk^\prime}}{\sqrt
{\xi^2_{Kk^\prime}+\Delta^2_{Kk^\prime}}}}.
\end{equation}
It is obvious that, in the case of repulsion between particles
composing $\bm K$--~pair, BCS~--~like solution, independent of $\bm
k $, is absent.


\section{ Superconducting gap}

The solutions of Eq.~(57) for SC energy gap in the cases of
attraction (${\tilde U}_K^{\ast}({\bm k}-{\bm k^\prime})<0$) and
repulsion (${\tilde U}_K^{\ast}({\bm k}-{\bm k^\prime})>0$)
between particles composing $\bm K$--pair differ from each other
essentially. First of all, we consider the case of attraction and
restrict ourselves to the simplest approximation Eq.~(43), namely,
${\tilde U}_K^{\ast}(k) \equiv U_K = \text {const}$. Such an
approximation, just as BCS one, enables one to get an explicit
expression for SC energy gap. A magnitude of the coupling constant
$U_K$ depends on the pairing mechanism which is not under
discussion here. The only circumstance we have to take into
account is that one may neglect a predominance of repulsion as
compared to attraction in comparatively narrow energy region
corresponding to a vicinity of zero-energy line of electron or
hole dispersion that is in the vicinity of the PFC. Let
$2\bar{\xi}$  be characteristic energy width of such a region and
suppose, for the sake of simplicity,  that $\bar{\xi}$ is more
less than any characteristic scale relating to each of the
subdomains $\Xi_K^{(-)}$ and $\Xi_K^{(+)}$, that is $\bar{\xi} <<
\varepsilon_{K-}$, $\bar{\xi} << \varepsilon_{K+}$. The
approximation ${\tilde U}_K^{\ast}(k) = \text {const}$ results in
that there is a solution of the Eq.~(57) independent of the
momentum of the relative motion of $\bm K$--pair that is
$\Delta_{Kk} \Rightarrow \Delta_K$. We restrict ourselves to a
consideration of such a solution only. Reducing, as usual, the sum
in Eq.~(57) into an integral over $\xi_{Kk}$ and introducing an
average density of states, $g_K$, related to unit area, one can
obtain the order parameter in the form
\begin{equation}
\label{58}\Delta_K\approx\bar{\xi}\cdot \exp \left (-{\frac
1{g_KU_K}} \right ).
\end{equation}
that is a solution which formally coincides with BCS solution. It
should be noted that $g_K$ is more less as compared with the total
density of states on the Fermi level. Therefore, in the case of
typical of phononic pairing mechanism coupling constant value,
one obtains the energy gap, Eq.~(58), which should be certainly
more less in comparison with the gap were arisen due to
conventional Cooper pairing on the full FC.

Now, let us consider the case when a repulsion between particles
composing $\bm K$--pair dominates. In this case, one has not to
take into account the existence of any bosonic degree of freedom
(phononic, electronic, magnetic or some else) as a necessary
condition of a rise of a bound state of $\bm K$--pair. Screened
Coulomb repulsion Eq.~(35) becomes the only essential interaction
with the effective coupling constant, Eq.~(43).

As one can see from Eq.~(57) there is no solution of constant
signs inside the domain $\Xi_K$ provided that $U_K>0$. Therefore,
to obtain an approximate solution of Eq.~(57) we suppose that the
order parameter dependence on the momentum of the relative motion
of $\bm K$--pair is given by a discontinuous function changing its
sign on the PFC. Restrict ourselves to a consideration of the
simplest case when the order parameter has independent of $\bm k$
and different from each other values inside the subdomains
$\Xi_K^{(-)}$ and $\Xi_K^{(+)}$. Namely, omitting the label $K$ in
the definition of the order parameter, we assume that $\Delta_K
\equiv \Delta_- >0$ inside $\Xi_K^{(-)}$ and $\Delta_K \equiv
-\Delta_+ <0$ inside $\Xi_K^{(+)}$. One can take note of the fact
that a singularity (a discontinuous character of the order
parameter in contrast to the solution of constant signs as it
follows from Eq.~(58) in the case of attraction) on the line
separating $\Xi_K^{(-)}$ and $\Xi_K^{(+)}$ may be considered as
one more manifestation of hyperbolic metrics of momentum space in
the same way as in the case of QSS when the wave functions,
Eqs.~(20), (21), of the relative motion of $\bm K$--pair, written
in the special case of the PFC position coinciding with the pair
center-of-mass energy, $2\varepsilon (\bm K /2)$, change their
behavior on the PFC as well.

The assumption relating to a character of the solution of Eq.~(57)
allows us to rewrite this equation in the form of the system of
two equations for $\Delta_-$ and $\Delta_+$. In this connection,
one should take into account the above mentioned remark that
effective interaction matrix element ${\tilde U}_K^{\ast} \sim
\Xi_K^{(-)}$ when both $\bm k$ and $\bm k^\prime$ belong to the
subdomain $\Xi_K^{(-)}$ that is a scattering due to the
interaction is restricted to this subdomain. However, if $\bm k$
belongs to $\Xi_K^{(-)}$ and $\bm k^\prime$ belongs to
$\Xi_K^{(+)}$ and hence the scattering is possible in the whole of
the domain $\Xi_K$, we have ${\tilde U}_K^{\ast} \sim \Xi_K$.
Taking account of that $\Xi_K << (2\pi)^2/a^2$, that is, in any
case, the scattering region is more less as compared with 2D
Brillouin zone, one may neglect a dependence of ${\tilde
U}_K^{\ast}({\bm k}-{\bm k^\prime})$ on the momenta $\bm k$ and
$\bm k^\prime$ except for that which is already taken into account
by the factor $\Xi_K$ in the definition of ${\tilde
U}_K^{\ast}({\bm k}-{\bm k^\prime})$. Assuming that ${\tilde
U}_K^{\ast}({\bm k}-{\bm k^\prime})\sim \Xi_K a^2$ where both $\bm
k$ and $\bm k^\prime$ belong to $\Xi_K$  and the coupling constant
is defined by Eq.~(43), one can rewrite Eq.~(57) in the form
\begin{eqnarray}
\label{59}
&& (1-\alpha)\Delta_- +\Delta_+ ={\frac
{U_K\Xi_Ka^2{h_{\alpha}}}{2S}} \Delta_-\sum_{k\in\Xi_K^{(-)}}
{\frac {f_k}{\sqrt{\xi_k^2 +\Delta^2_-}}} , \nonumber \\
&& \Delta_- +\alpha \Delta_+ ={\frac {U_K\Xi_Ka^2{h_{\alpha}}}{2S}}
\Delta_+\sum_{k\in\Xi_K^{(+)}} {\frac {f_k} {\sqrt{\xi_k^2
+\Delta^2_+}}}.
\end{eqnarray}
We denote here $h_{\alpha} \equiv (1-\alpha +{\alpha}^2)$ and
$\alpha \equiv {\Xi_K^{(-)}}/{\Xi_K}$. The factor $1-\alpha
+{\alpha}^2>0$ when $\alpha$ varies within the range $0\leq \alpha
\leq 1$. It arises as a difference between squared ``nondiagonal''
and a product of ``diagonal'' interaction matrix elements with
respect to the labels $(\pm)$ and depends on a filling of the
states inside the PFC. Thus, it takes into account statistical
correlations in the system of fermions composing $\bm K$--pairs.

Dependence of AF energy on the doping level allows us to hunt down
the evolution of the PFC due to a variation of doping. If $I(x) <
\delta \varepsilon_{KK^\prime}^{} \equiv I_m$, the PFC is absent
and $\varepsilon_{K-} = \varepsilon_0$ whereas $\varepsilon_{K+}
=0$ where $\varepsilon_0$ is an energy width of the domain
$\Xi_K$. At $I(x_2)=I_m$, there is an ``opening'' of the PFC at
two points, $a$ and $a^\prime$, which are situated on the
$k_1$--axis as it is shown in Fig.~2. A decrease of doping,
$x<x_2$, leads, first of all, to a rise, and then, to an extension
of the subdomain $\Xi_K^{(+)}$  which is accompanied with the
corresponding decrease of the subdomain $\Xi_K^{(-)}$. Thus, the
PFC length increases and then, after being up the maximal length
value corresponding to a certain doping level, becomes to
decrease, shrinking lastly at two points, $b$  and $b^\prime$, on
the $k_2$--axis (Fig.~2). Such a shrinking corresponds to a value
of doping level, relating to AF energy $I(x_1)=I_M$. If one assume
that the pair condensate density is directly connected to the PFC
length, one can qualitatively explain both rather small values of
the condensation density and a peculiar doping dependence of $T_C$
observed in cuprates. It should be noted that, generally speaking,
the choice of the domains $\Xi_K$ and $\Xi_{K^\prime}$ themselves
depends on doping as well since, in the last analysis, one has to
determine the antinodal vectors $\bm K$ and $\bm K^\prime$
minimizing the total electron energy. Later on, for the sake of
simplicity, we take into account such a dependence assuming that
it is explicitly included into AF energy $I(x)$ and screening
length $r_0(x)$ doping dependences. Such an evolution of the PFC
length with a variation of doping has to determine, in main,
doping dependence of both superconducting transition temperature
and superfluid density.

For the sake of simplicity, we suppose that $\varepsilon_{K-}$
and $\varepsilon_{K+}$ are linear functions of doping:
\begin{equation}
\label{60}\varepsilon_{K-}(x)=\varepsilon_0\cdot
{\frac{x-x_1}{x_2-x_1}}, \qquad\varepsilon_{K+}(x)=
\varepsilon_0\cdot {\frac{x_2-x}{x_2-x_1}}.
\end{equation}
Here, $x_1\leq x \leq x_2$. Let us introduce a ``reduced'' doping
level,
\begin{equation}
\label{61}y={\frac{x-x_1}{x_2-x_1}},
\end{equation}
varying within the limits of interval $0\leq y \leq 1$. Then we
have $\varepsilon_{K-}(y)=\varepsilon_0\cdot y$,
$\varepsilon_{K+}(y)=\varepsilon_0\cdot (1-y)$. As another one
simplification, we assume that the density of states is constant
inside the whole of the domain $\Xi_K$, $g_K =\Xi_K/(2\pi)^2
\varepsilon_0$. From this assumption, it follows immediately that
$\Xi_K^{(-)}/\Xi_K =\varepsilon_{K-}/\varepsilon_0$ that is
$\alpha \equiv y$.

Reducing the summation over momenta in Eq.~(59) to an integration
over the energy of the relative motion of $\bm K$--pair according
to
\begin{equation}
\label{62}\sum_{k\in\Xi_K^{(-)}} 1 \Rightarrow
Sg_K\int_{-\varepsilon_{K-}}^0  d\xi , \quad
\sum_{k\in\Xi_K^{(+)}} 1 \Rightarrow
Sg_K\int_0^{\varepsilon_{K+}}  d\xi
\end{equation}
one can rewrite, in the zero-temperature limit, the system of
equations, Eq.~(59), in the form
\begin{eqnarray}
\label{63}
&& (1-y)\delta_-+\delta_+ = w_K(y)\cdot h_y\cdot \delta_-
\cdot \ln{\left |{\frac
{y+\sqrt{y^2+\delta^2_-}}{\delta_-}} \right |}, \nonumber\\
&& \delta_-+y \delta_+ = \nonumber\\
&& = w_K(y)\cdot h_y\cdot \delta_+ \cdot \ln{\left
|{\frac {(1-y)+\sqrt{(1-y)^2+\delta^2_+}}{\delta_+}} \right |}
\end{eqnarray}
where $h_y\equiv(1-y+y^2)$, $\delta_{\pm} \equiv
\Delta_{\pm}/\varepsilon_0$ and
\begin{equation}
\label{64}w_K(y)={\frac {\pi e^2r_0(y)}{\varepsilon_0a^2}} \left
[{\frac {\Xi_Ka^2}{(2\pi)^2}} \right ].
\end{equation}
The screening length $r_0 =r_0(y)$ is, generally speaking, a
decreasing function of doping level. For the sake of simplicity,
we use a linear approximation for the coupling parameter,
Eq.~(64),
\begin{equation}
\label{65}w_K(y)=w_K\cdot \left (1-{\frac y{y_b}} \right ),
\end{equation}
where $w_K$ is determined by Eq.~(64) at $y=0$, and $y_b > 1$.

It should be noted that the chemical potential $\mu$ as a point of
reference of kinetic energy of the relative motion of $\bm K$--
pair changes due to a rise of SC order as compared with its value
in the normal (nonsuperconducting) state. However, the
corresponding shift of the chemical potential is quite small (of
the order of the SC gap). Thus, calculating the values of the
parameters $\delta_-$ and $\delta_+$ one need not consider the
chemical potential shift arising due to SC condensation of $\bm
K$--pairs. So, we can adopt approximately that $\mu$ is determined
by the only parameter $I$ and equals to the value which
corresponds to the PFC at given $I$ in normal state. However, in
contrast to BCS theory, in our case, just the chemical potential
shift determines, in main, the SC condensation energy and thus a
doping dependence of the superconducting transition temperature.

By definition, both unknown quantities $\delta_-$ and $\delta_+$
in Eq.~(63) are nonnegative: $\delta_-\geq 0$ , $\delta_+\geq 0$.
As one can easily see from Eq.~(63), the system of equations,
Eq.~(63), leads to the trivial solution $\delta_- = \delta_+ =0$
both at $y=0$ and $y=1$. As it also follows from Eq.~(63),
nontrivial solutions (if they exist) $\delta_-$ and $\delta_+$
coincide, $\delta_- = \delta_+$ at $y=0.5$. Nontrivial solutions
turn out to be possible under the condition that the coupling
parameter, Eq.~(64), is large enough. Dependence of this
parameter on doping, Eq.~(65), leads to an asymmetry of the
functions $\delta_-(y)$ and $\delta_+(y)$ that is, in general
case (except as some special values of doping level),
$\delta_-(y) \neq \delta_+(y)$. Doping dependence of
$\delta_-(y)$ and $\delta_+(y)$ is represented schematically in
Fig.~8.


\section{ Chemical potential shift}

In spatially homogeneous system, the value $2E_F$ of the chemical
potential of pairs indicates that the whole of the domain $\Xi_K$
is filled whereas all of the states inside any domain
$\Xi_{K^\prime}$ are vacant. A rise of spatially inhomogeneous
stripe structure leads to a hole redistribution between $\Xi_K$
and $\Xi_{K^\prime}$ with the result that the PFC arises. Thus the
full PFC may be treated, in the zero-temperature limit, as a line
separating filled and vacant pair states in momentum space. The
possibility of pairing itself resulting in an opening of the SC
gap on the PFC arises just as a result of such a redistribution
which may be due to above discussed partial restoration of AF
order. The number of vacant states inside $\Xi_K$ and, on the
other hand, filled states inside $\Xi_{K^\prime}$ is governed by
the value $I$ of AF energy which determines the position $2\mu$ of
the chemical potential of pairs with respect to the edges of the
energy bands corresponding to the domains $\Xi_K$ and
$\Xi_{K^\prime}$. The densities of states, $g_K(\varepsilon)$ and
$g_{K^\prime}(\varepsilon)$, corresponding to these domains are
represented in Fig.3. It is generally assumed \cite{35} that the
interactions between carriers which are not included into the
equivalent Hamiltonian of pairs such as Eq.~(45) do not affect
essentially the difference between the free energy values in the
N and SC states. Therefore, to calculate the chemical potential
shift due to the SC transition it is necessary to take into
account the Hamiltonian Eq.~(45) only. It should be noted that,
in conventional superconductors, the chemical potential shift is
equal to zero exactly because of the exact electron-hole
symmetry of the excitation spectrum \cite{64}.

In HTSC cuprates however, there is no sufficient reason for such a
statement \cite{47} just because of essential electron-hole
asymmetry \cite{11}. To evaluate the chemical potential shift
$\mu^\prime$ due to a condensation of $\bm K$--pairs belonging to
the domain $\Xi_K$ one has to take into account that a formal
definition of an average number of particles inside $\Xi_K$,
\begin{equation}
\label{66}\langle N_K^{}\rangle = 2\sum_{k\in\Xi_K} v^2_{Kk} +
\sum_{k\in\Xi_K} (u^2_{Kk} -v^2_{Kk})(n_{k,+1}^{} + n_{k,-1}^{}),
\end{equation}
takes into consideration the particles which may pass from $\Xi_K$
into $\Xi_{K^\prime}$. In thermal equilibrium, such a passage is
compensated by the particles passing from $\Xi_{K^\prime}$ into
$\Xi_K$. Therefore, the conserving quantity is a sum $\langle N_K
\rangle + \langle N_{K^\prime} \rangle$ where the second term is
an average number of particles inside $\Xi_{K^\prime}$. However,
the condensation may be considered in each of the domains $\Xi_K$
and $\Xi_{K^\prime}$ independently if, as it is accepted above,
$|{\bm K^\prime}-{\bm K}| >> \delta k_c$. In this case, one has to
take into account the only passages of particles across the PFC
($\Xi_K^{(-)} \leftrightarrows \Xi_K^{(+)}$, if one consider a
condensation of $\bm K$--pairs only) bearing in mind that the
position of the PFC in the normal state is determined by the AF
energy $I$ which is considered here as an external parameter. As
far as the chemical potential shift due to SC condensation of
pairs is also small together with $\Delta_{Kk}^{}$, such an
approximation only slightly affects the introduced below
coefficients $\lambda$ and $\tau$ and does not lead to any
qualitatively new results. Strictly speaking, the pairing
interaction, when $\Delta_{Kk}^{} \neq 0$, leads to, tending to
zero with $\Delta_{Kk}^{}$, a small change in the fluxes of
particles between the domains $\Xi_K$ and $\Xi_{K^\prime}$. Thus
$\langle N_K \rangle$ may be considered as approximately
conserving average number of particles inside the domain $\Xi_K$
provided that the AF energy $I$ has a certain given value. Taking
account of the explicit form of the coherence factor $v^2_{Kk}$,
Eq.~(55), one can calculate $\langle N_K \rangle$ in accordance
with Eq.~(66). In normal state, $v^2_{Kk} = 1$ when $k \in
\Xi_K^{(-)}$ and $v^2_{Kk} = 0$ when $k \in \Xi_K^{(+)}$ in
zero-temperature limit, therefore, the condition that $\langle N_K
\rangle = \text {const.}$ can be rewritten in the form \cite{4}
\begin{equation}\label{67}
\sum_{k\in\Xi_K^{(+)}} 1 - \sum_{k\in\Xi_K^{(-)}} 1 =
\sum_{k\in\Xi_K} {\frac
{\xi_{Kk}^{}}{\sqrt{\xi^2_{Kk}+\Delta^2_{Kk}}}}.
\end{equation}
First of all, let us consider the case of repulsion between
particles composing $\bm K$--pair. Reducing the summation over
momentum to integral over $\xi_{Kk}^{}$, after integration, one
obtains Eq.~(67) in the form
\begin{eqnarray}
\label{68}
&& \left [{\sqrt{(\varepsilon_{K+}-\mu^\prime)^2}
-\varepsilon_{K+}} \right ] - \left
[{\sqrt{(\varepsilon_{K-}+\mu^\prime)^2}-\varepsilon_{K-}} \right
] + \nonumber\\
 &&  + {\sqrt {{\mu^\prime}^2+\Delta^2_-}} - {\sqrt {{\mu^\prime}^2
+\Delta^2_+}} = 0.
\end{eqnarray}
As far as $\Delta_- << \varepsilon_{K-}$, $\Delta_+ <<
\varepsilon_{K+}$ in any case, and the chemical potential shift
$\mu^\prime$ measured from the PFC position at $\Delta
\rightarrow 0$ is small together with $\Delta$, one can
approximately rewrite Eq.~(68) as
\begin{equation}
\label{69}{\sqrt {{\mu^\prime}^2+\Delta^2_-}} - {\sqrt
{{\mu^\prime}^2 +\Delta^2_+}} \approx 2\mu^\prime + {\frac 12}
{\frac {\Delta^2_-}{\varepsilon_{K-}}} - {\frac 12} {\frac
{\Delta^2_+}{\varepsilon_{K+}}}.
\end{equation}
Assuming that $\mu^\prime = \mu_1^\prime +\mu_2^\prime$ where
$\mu_1^\prime$ ( $\mu_2^\prime$ ) is a quantity of the first
(second) order with respect to $\Delta$, the equation Eq.~(69)
may be solved with the use of the method of successive
approximations. Thus, finally we have
\begin{eqnarray}
\label{70}
&& \mu^\prime = {\frac {\varepsilon_0}{2\sqrt 2}}{\frac
{\delta^2_- - \delta^2_+}{\sqrt {\delta^2_- + \delta^2_+}}} +
\nonumber\\
&& + {\frac {\varepsilon_0}{16}} {\frac {(3\delta^2_- + \delta^2_+)
(\delta^2_- +3\delta^2_+)}{(\delta^2_- + \delta^2_+)}} \left
({\frac{\delta^2_+}{1-y}} - {\frac {\delta^2_-}{y}} \right ).
\end{eqnarray}
Two values of the order parameter, $\delta_-$ and $\delta_+$, are
not independent. They are connected with each other as it follows
from the equations Eq.~(63). For example, the first of these
equations allows us to express $\delta_+$ as a function of
$\delta_-$,
\begin{eqnarray}
\label{71}
&& \delta_+=\delta_- \cdot \gamma(\delta_-(y);y) , \qquad
\gamma(\delta_-(y);y)\equiv \nonumber\\
&& \equiv \left[w_K(y)h_y \ln \left |\frac
{y+\sqrt{y^2+\delta^2_-}}{\delta_-} \right |-(1-y) \right ].
\end{eqnarray}
The expression Eq.~(70) can be rewritten as
\begin{equation}
\label{72}\mu^\prime = \varepsilon_0\cdot\delta \cdot [\lambda +
\tau \delta]
\end{equation}
where we denote $\delta_- \equiv \delta$, and
\begin{eqnarray}\label{73}
&& \lambda \equiv {\frac 1{2\sqrt 2}}
{\frac {1-\gamma^2}{\sqrt {1+\gamma^2}}} , \nonumber\\
&& \tau\equiv{\frac 1{16}}\frac
{(3+\gamma^2)(1+3\gamma^2)}{(1+\gamma^2)^2} \left ( \frac
{\gamma^2}{1-y} - \frac 1y \right ).
\end{eqnarray}

In the case of attraction between particles composing $\bm
K$--pair, we have the only value of the order parameter, Eq.~(58),
which is independent of the momentum of the relative motion.
Therefore, to obtain the chemical potential shift due to SC
condensation one should formally write $\delta_+ = - \delta_-$.
Then we have
\begin{equation}
\label{74}\mu^\prime = - {\frac {\varepsilon_0}4}{\frac
{(1-2y)}{y(1-y)}}\delta^2.
\end{equation}
Thus, the approximation we use here leads to the absence of the
term which is linear in $\delta$ and the chemical potential shift
turns out to be proportional to $\delta^2$. In the symmetrical
case, $y=0.5$, when the PFC bisects the domain $\Xi_K$,
$\mu^\prime =0$; as it is seen from Eq.~(74), $\mu^\prime <0$
($\mu^\prime >0$) at $0<y<0.5$ ( $0.5<y<1$ ).

It should be noted that a necessity of $k$--dependence of the SC
gap and corresponding displacement of the chemical potential from
its value in the normal state was established phenomenologically
by Hirsch \cite{65} in his theory of hole superconductivity. It is
clear that ``the gap slope'' introduced by Hirsch is directly
related to our simple discontinuous solution of the gap equation
(6.13) whereas the linear term in the chemical potential shift,
arising just in the case when $\Delta_-\neq \Delta_+$, corresponds
to Hirsch's ``electron-hole symmetry-breaking term'' being the
difference, $\mu^\prime$, between the chemical potential values in
the superconducting and the normal state. One can see quite easily
that, in the case of electron-hole asymmetry observed in tunnel
current-bias characteristics, such a chemical potential shift is a
direct consequence of the particle conservation law. Indeed, if
one considers a redistribution of particles due to SC condensation
inside the domain $\Xi_K = \Xi_K^{(-)} + \Xi_K^{(+)}$ only it
becomes obvious that in the case when $\Delta_- \neq \Delta_+$,
the value $1/2$ of the coherence factor $v^2_{Kk}$, Eq.~(55),
cannot correspond to the position of the chemical potential
relating to the normal state (see Fig.~9). So, some chemical
potential shift is needed to satisfy the condition that the number
of transfers of holes from $\Xi_K^{(-)}$ must be equal to the
number of transfers into $\Xi_K^{(+)}$. The sign of such a shift
is determined by the sign of the difference $\Delta_- -\Delta_+$.
Thus, Hirsch's statement that $\mu^\prime>0$ is valid, generally
speaking, in the case when $\Delta_- >\Delta_+$. One can also note
that the so-called superconducting ``Fermi surface'' introduced in
\cite{65} as ``the locus in $k$-space of quasiparticle states of
minimum energy'', in a sense, plays the role which, indeed, plays
the PFC in the analysis of ARPES spectra and some other phenomena
typical of HTSC cuprates.


\section{ Condensation energy}

The existence of the solution of Eq.~(57) for SC order parameter
in doping interval $x_1<x<x_2$ does not mean yet that superfluid
SC state arises in the whole or, at least in some part, of this
interval. A phase transition from non-superfluid N state into
superfluid SC state occurs under a necessary condition: namely,
the condensation energy defined as a difference between N state
and SC state values of the ground state energy must be positive.

In the zero-temperature limit, a contribution into the
condensation energy which is associated with a condensation of
$\bm K$--pairs inside the domain $\Xi_K$ only, in accordance with
Eq.~(48), may be written as
\begin{eqnarray}
\label{75}
&& E_{0S}^{}=\sum_{k\in\Xi_K}\xi_{Kk}^{} - \nonumber\\
&& - \sum_{k\in\Xi_K} \left [ {\frac {\xi^2_{Kk}}{\sqrt
{{\xi^2_{Kk}}^2+\Delta^2_{Kk}}}} + {\frac 12}{\frac
{\Delta^2_{Kk}}{\sqrt {{\xi^2_{Kk}}^2+\Delta^2_{Kk}}}} \right ].
\end{eqnarray}
Reducing the summation over momentum in Eq.~(75) to the
integration over $\xi_{Kk}^{}$ one has to take into account that
the energy $\xi_{Kk}^{}$, Eq.~(49), measured from the chemical
potential of N phase varies within the interval
$-\varepsilon_{K-}^{}\leq\xi_{Kk}^{} \leq\varepsilon_{K+}^{}$.
Therefore, above discussed chemical potential shift due to SC
condensation, $\mu^\prime$, leads to
\begin{equation}
\label{76}\sum_{k\in\Xi_K}1\Rightarrow
Sg_K^{}\int_{-(\varepsilon_{K-}^{}+\mu^\prime)}^{-\mu^\prime} d\xi
+ Sg_K^{}\int_{-\mu^\prime}^{(\varepsilon_{K+}^{}-\mu^\prime)}
d\xi.
\end{equation}
Here we take into account the fact that SC order parameter has a
discontinuity on the PFC. Thus, correct to the terms of the order
of $\Delta^2$, Eq.~(75) may be represented in the form
\begin{equation}
\label{77}E_{0S}^{}=E_{0N}^{}-S\cdot 4g_K^{}\varepsilon_0^2\cdot
\delta \cdot ( \bar{\lambda} +c\delta ),
\end{equation}
where $\delta\equiv\Delta_-/\varepsilon_0$ is a dimensionless
order parameter and
\begin{equation}
\label{78}E_{0N}^{}=-S \cdot g_K^{} \varepsilon^2_K y^2
\end{equation}
is the corresponding contribution into the ground state energy of
the N phase. The parameters $\bar{\lambda}$ and $c$ are connected
with $\lambda$ and $\tau$ in Eq.~(73) in accordance with the
relations
\begin{equation}
\label{79}\bar{\lambda} = 2y\lambda ,\qquad {c=2y\tau +{\frac
{1+\gamma^2}{4}}}.
\end{equation}

The second term in the second expression in Eq.~(79) which is not
connected with the chemical potential shift due to SC condensation
may be formally related to a direct contribution of the pairing
interaction in the Hamiltonian Eq.~(45) into the condensation
energy whereas the contributions associated with the coefficients
$\lambda$ and $\tau$ may be related to a renormalization of
kinetic energy of the relative motion of $\bm K$--pairs being a
result of SC condensation. Although such a separation
\cite{46',47,48} of the condensation energy into kinetic and
potential energy contributions has an arbitrary character (it is
clear that both contributions vanish when the coupling constant in
Eq.~(45) tends to zero), it enables one to imagine more clear how
hyperbolic metrics of momentum space affects the ground state of
the electron system.

As it follows from Eq.~(77), an energy gain due to the
condensation of $\bm K$--pairs  is possible when
\begin{equation}
\label{80}\bar{\lambda}+c\delta >0.
\end{equation}
It is seen from Eq.~(77) that this gain is, in main, due to a
renormalization of the kinetic energy of the relative motion of
$\bm K$--pair. Indeed, the chemical potential shift due to a rise
of a condensate of $\bm K$--pairs results in the corresponding
shift of the position of the PFC. Provided that the condition
Eq.~(80) is satisfied the PFC is shifted in a way that there is an
extension of the part, $\Xi_K^{(-)}$, of the domain $\Xi_K$ in
which the energy of the relative motion of $\bm K$--pair is
negative. The ground state energy decreases just due to a filling
of the states which arise as a result of such PFC shift.

In this connection, one relevant optical experiment \cite{66}
consistent with the conception elaborating here should be noted.
An estimation of the superfluid density $\rho_s$ which is directly
connected with IR reflection, indicate that, in several HTSC
cuprates, $\rho_s$ significantly exceeds the value obtained from
optical conductivity by means of Kramers~-- Kronig relations under
the condition that one takes into account an energy interval
comparable to the SC gap \cite{66}. This contradiction may be
eliminated if one considerably extends the interval of
integration. In conventional superconductors, as it follows from
the BCS theory, each Cooper pair leads to an energy gain of the
order of $\Delta$. The energy width of the condensation region in
the vicinity of the FC is of the same order, $\sim\Delta$.
Therefore, the condensation energy turns out to be of the order of
$\Delta^2$. This explains the fact that, using the Kramers~--
Kronig relations, we can restrict ourselves to a finite interval
of integration having a character energy width of about $\Delta$.
The presence of the linear term ($\sim\Delta$) in the condensation
energy in Eq.~(77) clearly indicate that each $\bm K$--pair also
leads to an energy gain of the order of $\Delta$ but this gain
must be associated with the condensation region in momentum space,
which is connected not with the FC but with the PFC, having a
character energy width of about a character kinetic energy of the
relative motion inside the subdomain $\Xi_K^{(-)}$, namely,
$\varepsilon_{K-}$, which is usually more large as compared to
$\Delta$.

Let us define the condensation energy per unit area as
\begin{equation}
\label{81}\varepsilon_c\equiv
(E_{0N}^{}-E_{0S}^{})/4g_K^{}\varepsilon_0^2S = \delta\cdot
(\bar{\lambda}+c\delta)
\end{equation}
and study qualitatively its dependence on doping level. In
Fig.10, we represent a plot of the function Eq.~(81) calculated
numerically for some values of the coupling parameter, Eq.~(64).
It is obvious that there exists a certain minimal value of this
parameter which corresponds to a start of SC condensation. This
conclusion is in agreement with a finite value of the scattering
amplitude at $\omega=\omega_1$ obtained in Sec.~IV.

As one can see from Fig.~10, calculated condensation energy has
negative sign inside some region of doping level. This fact is,
mainly, due to negative sign of the chemical potential shift
$\mu^\prime$ leading to kinetic energy increase. One may believe
that such a result is a consequence of a special choice of the gap
equation solution being discontinuous on the PFC and leading to an
energy gain in a doping region where the condensation energy turns
out to be positive. At another choice of the gap equation
solution, which varies with a momentum of the relative motion
continuously within an energy scale of the order of $\Delta$ near
the PFC inside both $\Xi_K^{(-)}$ and $\Xi_K^{(+)}$, the values of
$\Delta_-$ and $\Delta_+$ are supposed to be unaffected ,
therefore, one may expect a gain in the condensation energy in the
whole of the reduced doping interval, $0<y<1$, in which the
solution of the equation (6.13) exists. In this connection, it
should be noted that such a choice of the parameters $\Delta_-$
and $\Delta_+$ has to correspond to more symmetric tunnel
current-bias characteristics in extremely underdoped regime in
comparison with the optimal one.


\section{ Superconductivity at weak ferromagnetic ordering}

It is known \cite{67,68} that, due to doping, charge-density wave
(CDW) may arise in AF phase resulting from spin-density wave
(SDW). A coexistence of SDW and CDW may lead to a rise of weak
ferromagnetic state as it is, for example, in a boride family
\cite{69}. Charge ordering is also known in doped superconducting
cuprates containing bismuth \cite{70}. 1D stripe structure of
such cuprates leads naturally to a modulation of spin and charge
density and, as a result, makes possible a rise of weak
ferromagnetism. In this Section, using the conception of PFC, we
briefly discuss a possibility of superconductivity in cuprates
with weak ferromagnetic ordering in AF parts of stripes \cite{7}.

If one suppose that a value of corresponding spontaneous
magnetization is large enough, $\mu_B H_e >> \Delta$ (here,
$\mu_B$ is the Bohr magneton, $H_e$ is ferromagnetic Weiss field,
$\Delta$ is SC gap in the absence of the ferromagnetic order), the
only trivial solution for SC order parameter may be obtained when
one considers the domains $\Xi_K$ and $\Xi_{K^\prime}$
independently from each other. Indeed, when Weiss field is large
enough, the subbands corresponding to spin quantum numbers of
opposite sign become removed with respect to each other so that
the domain of definition of a momentum of $\bm K$--pair relative
motion turns out to be empty. Such a conclusion is consistent with
that which follows from the problem of Cooper pairing in a weak
ferromagnetic studied in \cite{9,10}. A similarity between Cooper
pairing and $\bm K$--pairing becomes clear if one takes into
account the fact that, in the case of Cooper pairing, the momenta
of particles composing a pair play role of the momenta of the
relative motion as well. Thus, to obtain a nontrivial solution for
SC order parameter one needs in a consideration of above mentioned
electron and hole transfers across the parts of the PFC, one of
which is related to AF part of a stripe and belongs to the domain
$\Xi_K$ and another one corresponds to M part of a stripe
associated with the domain $\Xi_{K^\prime}$. Therefore, in the
following we take into account the transfers $\Xi_K
\leftrightarrows \Xi_{K^\prime}$ only. Note that in \cite{71'} the
model with strong tendency to spatial modulation due to
correlations between M and AF parts of a stripe structure is
considered in connection with the problem of high-temperature
superconductivity in cuprates.

To obtain the Hamiltonian of hole pairs in the case discussed
here, one needs in taking account of the fact that each hole
passing from $\Xi_{K^\prime}^{(-)}$ into $\Xi_K^{(+)}$ in momentum
space passes from M part into AF part of a stripe in real space.
As a result, this hole arises in a region with weak ferromagnetic
ordering due to a coexistence of SDW and CDW. It seems quite
natural to propose that an average magnetization in this region
is proportional to above introduced phenomenological parameter $I$
which may be considered, in a certain sense, as AF Weiss field,
$\mu_B H_e = \chi I$. A dimensionless parameter $\chi$ has to be
considered as a small quantity, however, as it is noted above, we
suppose that $\Delta/I << \chi <<1$. One can write a hole energy
as
\begin{equation}
\label{82} \varepsilon_{{\sigma}{{\bm k}_{\pm}}}^{} \equiv
\varepsilon_{\sigma}^{} ({\bm k}_{\pm}) = \varepsilon({\bm
k}_{\pm}) + \chi I\sigma\cdot\Theta_k,
\end{equation}
Here, $\Theta_k$ is a characteristic function such that $\Theta_k
= 1$ if a hole momentum belongs to $\Xi_K$, and $\Theta_k = 0$ in
any other case. Thus, the Hamiltonian of pairs corresponding to
Eq.~(45) has the form
\begin{eqnarray}\label{83}
\hat H_K^{}= && \sum_k \left [(\varepsilon_{{\uparrow} {{\bm
k}_+}}-\mu) {\hat{a}^{\dag}_{k+}}{\hat{a}^{}_{k+}} +
(\varepsilon_{{\downarrow} {{\bm k}_-}})-\mu)
{\hat{a}^{\dag}_{k-}}
{\hat{a}}^{}_{k-}\right .] \nonumber\\
&& +{\frac 1{S}} \sum_{k,k^\prime} {\tilde U}_K^{\ast}({\bm k}- {\bm
k^\prime}) {\hat{a}^{\dag}_{k+}} {\hat{a}^{\dag}_{k-}}
{\hat{a}^{}_{{k^\prime}-}} {\hat{a}^{}_{{k^\prime}+}}
\end{eqnarray}
here ${\hat{a}_{k_+\uparrow}^{}}\equiv {\hat{a}_{k+}^{}}$ and
${\hat{a}_{k_-\downarrow}}\equiv {\hat{a}_{k-}}$. The summation
over $\bm k$ and $\bm k^\prime$ is taken over all range of momenta
of the relative motion of $\bm K$ and $\bm K^\prime$--pairs. As it
is well known \cite{71}, spin-dependent kinetic energy in the
Hamiltonian of pairs results in a redefinition of quasi-particle
energies: instead of Eq.~(51) we have
\begin{equation}
\label{84}\eta_{K\beta}^{}({\bm k}) = {\sqrt {\xi^2_{Kk} +
\Delta^2_{Kk}}}+{\frac {\beta}2}{{\chi}I}\cdot\Theta_k,
\end{equation}
where $\beta = \pm 1$. One can see that $\eta_{+1}^{}(k) \geq 0$
at any $\bm k$ inside the domain $\Xi_{K^\prime}$ and therefore
the corresponding factor $f_k = 1$ in the zero-temperature limit.
On the other hand, inside the domain $\Xi_K$, the condition that
$\eta_{-1}(k) \geq 0$ may be satisfied if only
\begin{equation}
\label{85}{\chi}I{\sqrt {1-({2\Delta_{Kk}}/{\chi}I)^2}}\leq
2\xi_{Kk}^{} \leq2\varepsilon_{K+}^{},
\end{equation}
where $\varepsilon_{K+}$ is an energy width of the subdomain
$\Xi_K^{(+)}$. Under the condition Eq.~(85), the factor $f_k^{} =
1$ whereas in the opposite case when
\begin{equation}\label{86}
0 \leq 2|\xi_{Kk}^{}| \leq {\chi}I{\sqrt
{1-({2\Delta_{Kk}^{}}/{\chi}I)^2}},
\end{equation}
this factor is equal to zero because of the equalities $n_{k,+1}
= 0$, $n_{k,-1} = 1$ in the zero-temperature limit. Thus, the
factor $f_k$ excludes some part of the domain $\Xi_K$ from the sum
in the equations Eqs.~(56), (57) which determine SC order
parameter. It should be noted that the condition $\Delta << \chi
I$ allows us to simplify Eqs.~(85), (86) as far as $\sqrt
{1-(2\Delta/\chi I)^2} \approx 1$.

Let us consider the simplest solution of Eq.~(57) when
$\Delta_{Kk} \equiv \Delta_- >0$ inside the subdomain
$\Xi_{K^\prime}^{(-)}$ and $\Delta_{Kk} \equiv -\Delta_+ <0$
inside the subdomain $\Xi_K^{(+)}$. Then, one can reduce the
summation over momenta in Eq.~(59) to an integration over the
energy of the relative motion of pairs according to
\begin{equation}
\label{87}\sum_{k\in\Xi_{K^\prime}^{(-)}} 1 \Rightarrow
Sg_{K^\prime}^{}\int_{-\varepsilon_{{K^\prime}-}^{}}^0 d\xi, \quad
\sum_{k\in\Xi_K^{(+)}} 1 \Rightarrow
Sg_K^{}\int_{{{\chi}I}/2}^{\varepsilon_{K+}^{}} d\xi
\end{equation}
where $\varepsilon_{{K^\prime}-}^{}$ is an energy width of the
subdomain $\Xi_{K^\prime}^{(+)}$. Let us assume, for the sake of
simplicity, that $\varepsilon_{{K^\prime}-}^{} =
\varepsilon_{K+}^{}$; then average densities of states, $g_K^{}$
and $g_{K^\prime}^{}$, are equal to each other, $g_{K^\prime}^{} =
g_K^{}$. Also, we assume that $\Xi_{K^\prime}^{(-)} =
\Xi_K^{(+)}$, then we have simply $\alpha=0.5$. With the help of
Eq.~(87), the equations Eq.~(59) may be rewritten as
\begin{eqnarray}\label{88}
\frac {\delta_-} {2} + \delta_+ =\bar{w}_K\cdot \left (1-{\frac
y{y_b}} \right ) (1-y) \delta_- \cdot \ln{\frac
{2(1-y)}{\delta_-}} , \nonumber\\
\delta_- + \frac{\delta_+}{2}  = \bar{w}_K\cdot \left (1-{\frac
y{y_b}} \right ) (1-y) \delta_+ \cdot \ln{\frac
{2\varepsilon_0(1-y)}{{\chi}I}}
\end{eqnarray}
where $\bar{w}_K = 3w_K/4$. The second of the equations Eq.~(88)
is linear, therefore,
\begin{equation}
\label{89}\delta_+=\gamma \delta_-,
\end{equation}
where, as before, $\delta_{\pm} =\Delta_{\pm}/\varepsilon_0^{}$,
$\varepsilon_{K+}^{} =\varepsilon_0^{}\cdot (1-y)$,
\begin{equation}
\label{90}\gamma = \frac 1{\bar{w}_K\cdot \left (1-{\frac y{y_b}}
\right ) (1-y) \ln\frac {2{\varepsilon}_0(1-y)}{{\chi}I} - \frac
12}.
\end{equation}
As far as $\gamma\leq0$ by definition and on account of Eq.~(3),
the solution of the equations Eq.~(88) exists within a bounded
range of AF energy values,
\begin{equation}
\label{91}\delta \varepsilon_{KK^\prime}^{} < I <{\frac
{2\varepsilon_0(1-y)}{\chi}} \exp \left [-{\frac
1{2\bar{w}_K}}{\frac {y_b}{(1-y)(y_b-y)}} \right ].
\end{equation}
Thus, the inequalities Eq.~(91) determine that range of doping in
which the SC order parameter may be unequal to zero.

The first of the equations Eq.~(88) leads to dependent on doping
absolute value of the SC order parameter,
\begin{equation}
\label{92}\delta_-=2(1-y)\cdot \exp \left [-{\frac
1{2\bar{w}_K}}{\frac {y_b(1+2\gamma)}{(1-y)(y_b-y)}} \right ].
\end{equation}
In spite of the fact that the expression Eq.~(92) is formally
similar to BCS gap, Eq.~(92) depends on the coupling constant
$w_K$ in essentially complicated way because the parameter
$\gamma$, Eq.~(90), depends on $w_K$ itself. The pre-exponential
factor in Eq.~(92) is limited by a character value of kinetic
energy of the relative motion of $\bm K$--pair inside the
subdomain $\Xi_K^{(+)}$. This energy itself depends on that to
what extent the domain $\Xi_K$  is filled by carriers or, in other
words, to what extent the PFC is opened due to a rise of a stripe
structure.

Considering hole number conservation inside the domain
$\Xi_{K^\prime}$, one can obtain the chemical potential shift in
the form of Eqs.~(72), (73) where the parameter $\gamma$ is
defined by Eq.~(90). In Fig. 11, doping dependence of the
condensation energy, Eq.~(81), and the gap parameter, Eq.~(92),
are represented for the value of the coupling constant $w_K=1$.
It should be noted that, in the case discussed here, this
constant has to be considered as a reduced value of the parameter
defined by Eq.~(43) because of a spatial separation of carriers
composing a pair.

As one can see, the SC energy gap takes finite values at certain
doping levels corresponding to zero condensation energy. As far as
the SC transition temperature $T_C$ is directly connected with the
condensation energy, it is obvious that, in contrast to well-known
consequence of the BCS theory, namely, $2\Delta/T_C\approx3.5$,
any universal, independent on doping, relation between $\Delta$
and $T_C$ is absent. Therefore, as it follows from the
consideration of doping dependencies of SC gap and condensation
energy presented in Sections IX and X, one must not consider as
striking the large values of the ratio $2\Delta/T_C$ observable in
underdoped HTSC cuprates \cite{72}.


\section{ Summary and conclusions}

Specific quasi-two-dimensional electron structure of HTSC cuprate
compounds results from their layered crystal structure and
chemical composition. Namely, if carrier concentration is close
to the half-filling, the considerable part of the hole Fermi
contour with strong nesting features turns out to be belonging to
an extended region of momentum space with hyperbolic metrics.
Such an electron structure leads to a possibility of a rise of
electron and hole pairs with large, of the order of doubled Fermi
momentum, total momentum and comparatively small momentum of the
relative motion of the pair \cite{1,2}. Due to hyperbolic metrics
of momentum space bounded states may exist both in the case of
attraction and repulsion between particles composing the pair. As
it follows from the qualitative analysis of the energy dependence
of the scattering amplitude, in both cases (repulsion and
attraction), the scattering amplitude has two poles. In the case
of repulsion, one of the poles, with larger positive energy and
infinitesimal positive decay, may be associated with a
quasi-stationary state of the relative motion of the pair
\cite{1,2} whereas the second pole, with smaller positive energy
and finite negative decay, may lead to a development of
superconducting instability and the imaginary part of this pole
may be related to the SC gap. An evaluation of Coulomb repulsion
in the electron system with and without QSS's indicates that,
provided that QSS concentration exceeds a certain value, an
energy gain is possible and thus incoherent QSS's may exist
resulting in some suppression of one-particle density of states
\cite{2}. Therefore, a rise of such QSS's may be directly related
to the pseudogap state being one of the remarkable features of
underdoped HTSC cuprates. It is obvious that positive real part
of any of the two poles results in an energy increase when the
pairs arise and this increase might hardly be compensated only
due to a rise of the SC condensate or sufficiently large QSS
concentration. One can assume that such an energy increase is
associated with the fact that, to give rise to a creation of the
pairs, at least a part of the domain of definition of a momentum
of the relative motion of the pair were free of carriers. To set
this part free one must annihilate the particles composing the
pair inside the FC and then create them in a new pair state
outside of the FC. Such a transfer of a number of pairs in
momentum space may result in a redistribution of carriers in real
space. Total energy increase which is due to this redistribution
may be compensated when, as it appears to be just the case being
related to hole-doped HTSC cuprates, there is an energy
decrease due to partial restoration of AF ordering in
hole-depleted regions arisen. An alternation of hole-depleted and
enriched regions in real space forms charge and spin spatial
structure (irregular and dynamic, generally speaking) of the
electron system. Under definite conditions (in particular, in
underdoped regime), this structure, associated closely with
short-range AF order fluctuations, becomes apparent as quasi-regular
static or dynamic 1D stripe structure. Such phase
separation \cite{73} accompanied with a change of a filling of
hole states in momentum space may correspond to the minimum of
total energy of the electron system. Short-range AF ordering
stabilizing stripes due to redistribution of hole pairs is
intrinsic but not unique possible attribute of such a
self-organization. For example, long-range orbital magnetic ordering
\cite{74',74,75,75'}, known as a flux phase state, may play the same role
as well. Although there is not any direct evidence of existence
of flux phase state in cuprates, this phase perhaps is associated
with the so-called hidden-order-parameter region in the
phase diagram \cite{76}.

Such a scenario of total energy gain due to carrier redistribution
is not the only one. Another one possibility, analogous to that
which may be in the case of the superconductivity \cite{77},
may be related to the problem here discussed. Namely, one may to consider
an increase of AF transition temperature resulting from a change of an
excitation energy distribution  due to
transitions from $\Xi_K^{(+)}$ into $\Xi_{K^\prime}^{(-)}$.

In any of the cases considered here, a redistribution of carriers
in momentum space may result in a rise of new
zero-excitation-energy line separating occupied and vacant states in 2D
Brillouin zone. We believe that, first of all, a rise of vacant
states inside and occupied states outside of the FC must lead to
a formation of pairs with total momenta corresponding to the
largest areas $\Xi_K$ and $\Xi_{K^\prime}$ with ${\bm K}$ and
${\bm K}^\prime$ along the antinodal directions. Such pairs have
the largest binding energies and exist up to the temperature,
$T^{\ast}$, of the beginning of the pseudogap regime. Then, the
lowering of the temperature from $T^{\ast}$ to $T_C$ results in a
gradual rise of electron and hole pairs with total momenta having
different values and directions and corresponding to some set of
domains $\Xi_K$ and $\Xi_{K^\prime}$. At last, at $T = T_C$, there
is a beginning of SC condensation of pairs into the state with
the largest binding energy. SC condensation gives a start to the
growth of the unoccupied part, $\Xi^{(+)}_K$, of the
``antinodal'' domain $\Xi_K$ at the expense of a redistribution
of carriers both inside and outside of the FC. Thus, as a final
result, zero-energy line for pair excitations arises inside the
domain $\Xi_K$ (similarly, such a line arises inside the
corresponding domain $\Xi_{K^\prime}$ outside of the FC as well).
Just this line may be treated as ``pair'' Fermi contour (PFC).

The conception of PFC and hyperbolic pairing enables one to
explain qualitatively some general features of phase diagram and
many surprising experimental data relating to high--$T_C$
cuprates. In particular, a rise of both SC and pseudogap state
may be considered as a manifestation of hyperbolic metrics of
momentum space and screened Coulomb repulsion between holes.
Therefore, both SC gap and pseudogap must have one and the same
energy scale and their $d$--type ``orbital'' symmetry, in fact, is
determined by the crystal symmetry. A character spatial scale of
a pair both in QSS and SC state (the coherence length) is of the
order of a few interatomic distances \cite{2}. As an evidence in
favor of the PFC conception, one may consider an interpretation
of two interesting experiments, relating to examination of
electronic spectrum of several of HTSC cuprates with the help of
ARPES technique. In the case, when an energy of excited electron
is near the Fermi level $E_F$ (less than about 100 meV),
ARPES data \cite{78} indicate unequivocally that electronic
structure has 2D character and FC remains in the well-known form
of a square with rounded corners \cite{11}. If an electron is
excited far from $E_F$ (about 500 meV) the electronic
structure becomes rather 1D than 2D. It should be noted that,
truly, a simple cross-shaped form of the FC, in fact, derived not
directly from experimental data but offered as a result of
motivated speculations based on the simplest 1D stripe model.
Thus, the ARPES data \cite{78} evidence the simultaneous
existence of both FC and PFC.

Such fairly surprising conclusion is entirely consistent with the
concept of the PFC introduced here. In fact, to detect the PFC
using ARPES technique one needs an electron excitation with an
energy which is essentially less than $E_F$: if excited state is
situated near the PFC, the excited electron may easily find a
partner to form a pair.

One can believe that the so-called ``dip-hump structure'' in
the ARPES spectra \cite{79} is one more evidence in favour of the
concept of PFC: a hump, arising (just in the case of antinodal
direction corresponding to the maximal value of $\Xi_K$) at
energies which are essentially more than the energy related to
the quasi-particle peak, may be connected with the excitations
in the form of pairs near the PFC.

Another one unusual feature of HTSC cuprates with $d$--type
symmetry of the SC gap can be qualitatively interpreted in the
framework of the PFC conception. It is believed that impurity
scattering has to lead to essential reducing of $T_C$ because the
scattering of a pair into regions of momentum space corresponding
to nodal directions means, in fact, a break of the pair. In this
sense, nonmagnetic impurities play role similar to that which
play magnetic impurities in conventional superconductors.
However, the experimental fact is that the HTSC cuprates are
weakly sensitive to impurity content. Thus, there are no
experimental data which were confirmed such a ``destructive''
influence of impurities on the SC state. In this connection, it
should be noted that, to reconcile such a statement with the
experiment, one has to suppose that any interaction resulting in
a scattering of Cooper pair in $d$-type superconductor, including
the interaction which leads to a binding in the pair itself, must
possess the peculiar feature, namely, the scattering into ``nodal
regions'' has to be more weak in comparison with the scattering
into ``antinodal regions'' (the so-called ``forward scattering'')
\cite{80,81,82}. Using the PFC conception, it is not necessary to
consider any peculiar feature of a scattering as far as the
scattering inside the domain $\Xi_K$, that is a variation of the
direction of ${\bf K}$--pair relative motion momentum, is, in
fact, almost isotropic as in the case of $s$--scattering of Cooper
pairs in conventional superconductors.

A problem connected with a strong anisotropy of reversed
relaxation times, that is an existence of the so-called ``hot''
and ``cold'' spots on the Fermi surface \cite{80,81,82,83}, can be
qualitatively solved in the framework of the PFC conception too.
Indeed, a rise of ${\bm K}$--pairs results in their free in-plane
motion without a change of charge density whereas a character of
the interaction of paired carriers may be changed essentially:
this interaction, being inside antinodal regions, turns out to be
more weak as compared to the interaction of unpaired carriers
inside nodal regions \cite{2}.

The idea we use here is based, in main, on the fact that the PFC
should be ``opened'' that is, due to a rise of stripes and
hyperbolic metrics of momentum space, some piece of the FC turns
out to be the same as a line of zero kinetic energy of the
relative motion of a hole pair with large momentum. If such a line
corresponding to a certain ${\bm K}$ is close enough to rather
large piece of the real FC (such a case may occur, for example,
just at ${\nu <<1}$ and ${|2{k_F}-K|<<k_F}$) the pairing mechanism
here discussed may be possible as well even without any hole
redistribution both in momentum space (between the domains
${\Xi}_K$ and $\Xi_{K^\prime}$) and in real space (that is without
a rise of a stripe structure). In this case, the value of
${\varepsilon}_{K+}$ plays role of a cut-off parameter since it
must appear in the arguments of the logarithmic functions in
Eq.~(63) together with the SC gap parameters ${\Delta}_{-}$ and
${\Delta}_{+}$. In a sense, the pairing problem becomes analogous
to that which arises in the case of Cooper pairing in weak
ferromagnets \cite{9,10}. This statement is consistent with the
results presented in Sections~III and IV. Namely, at
${\nu}{\rightarrow}0$ and ${{\omega}_1}{\rightarrow}0$ (it is
obvious that the parameter ${{\omega}_1}{\rightarrow}0$ plays role
of an energy distance between the FC and the line of zero energy
of pair relative motion), the imaginary part of the SC pole of the
scattering amplitude (which is proportional to the SC gap
parameter) exceeds necessarily the real part of the pole due to a
logarithmic singularity of $B_{K2}(\omega)$ at ${\omega}\to 0$, as
one can see from Fig.~5. Thus, one can assume a possibility of the
existence of crystals with such a form of the FC which is
optimally conforming with the form and energy position of a line
of zero kinetic energy of the relative motion of a hole pair with
large total momentum (one can consider hyperbolic lines used here
as a certain limiting case). As a result, in such crystals, the
pairing mechanism discussed here may dominate (possibly, even
without AF state in the neighborhood of the SC one and thus
without a rise of stripes as it maybe occurs in cuprate compounds
with more than one $CuO_2$ plane in the unit cell \cite{83'} ).

It should be noted that a superconducting state with large ($K
\approx 2k_F$) total pair momentum was previously studied
\cite{84} in the framework of the microscopic model of the
coexistence of superconductivity and antiferromagnetism or
charge-density wave (a structural phase transition; in such a
case, the momentum $K \approx 2k_F$ turns into a new vector of
the reciprocal lattice). In this model, the state with large pair
momentum arises as a result of coexistence of AF ordering and
Cooper pairs with zero total momentum. Phenomenologically, such
transitions as a break of corresponding symmetry is usually
considered in the framework of, for example, SO(5) or SU(4)
models \cite{85,86}. Zhang et al. \cite{85} have assembled AF and
$d$--wave SC order parameters into a five-dimensional vector
and have postulated the symmetry of unified in such a way order
parameter under rotations of an SO(5) group. However, to obtain
the closed Lie algebra of fermion pairing and particle-hole
operators describing antiferromagnetism and superconductivity,
one has to consider a more general, than SO(5) symmetry, for
example, the SU(4) symmetry \cite{86}. Such more general approach
leads directly to a rise of the components of the unified order
parameter corresponding to pairs with large, of the order of the
AF vector, total momentum \cite{86}. Thus, one may conclude that
$\bm K$ and $\bm K^\prime$--~pairs introduced in our paper in a
microscopic way are fully consistent with rather general symmetry
constraints. Note that if the vector ${\bm K}$ coincides with the
AF vector exactly the SC order parameter due to a rise of ${\bm
K}$--~pairs and the AF (triplet) order parameter turn out to be
connected to another one SC order parameter corresponding to
pairs with zero total momentum. A small difference between ${\bm
K}$ and the AF vector leads to a small total momentum of these
pairs. Such pairs may be in singlet spin state (conventional
Cooper pairs) or in triplet spin state. The case we discuss in
this paper corresponds just to the latter of the two
possibilities, namely, triplet AF order coexists with singlet SC
order due to ${\bm K}$--~pairs and triplet SC order due to the
pairs with small total momentum.

The phenomenological approach used here to take account of the
influence of AF fluctuations on carrier pairing enables one to
interpret qualitatively the key experimental data relating to HTSC
cuprates. We believe that the principal conception of hyperbolic
pairing and a rise of pair Fermi contour is an inherent feature of
cuprate electron system which has to become apparent both in band
scheme and in appropriate models of strong-correlated systems,
such as ${t-J}$ model \cite{87} with regard for
next-nearest-neighbor interactions (the so-called
${t-{t^\prime}-J}$ model \cite{88}), description of the electronic
structure.

\begin{acknowledgments}

We are deeply grateful to A.~F.~Andreev, A.~M.~Dykhne, V.~L.~Ginzburg,
Yu.~Kagan, L.~V.~Keldysh and Yu.~E.~Lozovik for fruitful discussions.
The work was supported, in part, the Russian
scientific-educational program ``Integration''
(projects AO133 and AO155).
\end{acknowledgments}


\newpage

\begin{figure}
\caption{\label{fig:epsart} Phase diagram (temperature vs doping
level) typical of hole-doped HTSC cuprates.}
\end{figure}

\begin{figure}
\caption{\label{fig:epsart} Typical of hole-doped HTSC cuprates,
hole Fermi contour (FC) as a square with rounded corners (labeled
by the Fermi energy, $E_F$) centered at ($\pi , \pi$)  . The
domain of definition of momenta of the relative motion of $\bf K$
and $\bf K^\prime$-- pairs are denoted as $\Xi_K$ and
$\Xi_{K^\prime}$, respectively. Each such a domain consists of two
parts, $\Xi_K^{(-)}$, $\Xi_K^{(+)}$ and $\Xi_{K^\prime}^{(-)}$,
$\Xi_{K^\prime}^{(+)}$, respectively. Inside the subdomains
$\Xi_K^{(-)}$, $\Xi_{K^\prime}^{(-)}$ ($\Xi_K^{(+)}$,
$\Xi_{K^\prime}^{(+)}$), the energy of the relative motion of
corresponding pair measured from the pair chemical potential value
$2\mu$ is negative (positive). Total pair momentum is directed
along an antinodal direction. The lines separating the subdomains
of negative and positive relative motion energy form the pair
Fermi contour (PFC). Doping decrease results in an opening of the
PFC at two points, $a$ and $a^\prime$, on $k_1$--~axis,
corresponding to a doping level $x_2$. Then, there is a rise and
an extension of the subdomains $\Xi_K^{(+)}$ and
$\Xi_{K^\prime}^{(-)}$ accompanied with the corresponding decrease
of the subdomains $\Xi_K^{(-)}$ and $\Xi_{K^\prime}^{(+)}$. The
PFC shrinks at two points, $b$ and $b^\prime$, on $k_2$--~axis,
corresponding to a doping level $x_1 < x_2$.}
\end{figure}

\begin{figure}
\caption{\label{fig:epsart} Top panel: a sketch of the domains
$\Xi_K$, $\Xi_{K^\prime}$ and hole distribution in the cases
corresponding to a homogeneous state of the electron system (left
top panel) and a stripe state (AF part of a stripe, middle top
panel; M part of a stripe, right top panel). Occupied and
unoccupied pair states are separated by the PFC. Occupied states
inside the domains are shadowed. Bottom panel: relative-motion
band diagram for homogeneous state (left bottom panel), AF part of
a stripe (middle bottom panel) and M part of a stripe (right
bottom panel).}
\end{figure}

\begin{figure}
\caption{\label{fig:epsart} A sketch of a stripe ordering in
momentum space (top panel) and real space (bottom panel). Arrows
are hole transitions between AF and M parts of stripes.}
\end{figure}
\begin{figure}
\caption{\label{fig:epsart} A plot of the function
$B_{K2}(\omega)$, Eq. (4.2), schematically.}
\end{figure}
\begin{figure}
\caption{\label{fig:epsart}
Plots of the function $B_{K1}(\omega)$
: solid line - Eq.~(23), dashed line - Eq.~(30). An illustration
of the graphic solution of the equation
Eq.~(14), schematically.
}
\end{figure}
\begin{figure}
\caption{\label{fig:epsart} A comparison of typical of HTSC
cuprates phase diagram and the graphic solution of the equation
Eq.~(14), determining the poles of the scattering amplitude
(schematically).}
\end{figure}
\begin{figure}
\caption{\label{fig:epsart} Solutions, $\delta_-$ and $\delta_+$,
of the system of equations, Eq.~(63), plotted schematically as
functions of reduced doping level. Solid (dashed) line corresponds
to the value of the effective coupling constant $w_K=2$
($w_K=4/3$).}
\end{figure}
\begin{figure}
\caption{\label{fig:epsart} Solid line:
energy dependence of the coherence factor $v^2_{Kk}$ in the case
when $\Delta_- > \Delta_+$ (schematically). Dashed line
corresponds to $v^2_{Kk}$ without ane chemical potential shift
$\mu^\prime$. One can see that such a shift is a direct
consequence of the particle number conservation inside the domain
$\Xi_K$: this number in normal state (in this Figure: the
rectangular area corresponding to the subdomain $\Xi_K^{(-)}$) has
to be equal to the area bounded by the solid line.}
\end{figure}
\begin{figure}
\caption{\label{fig:epsart} Condensation energy, Eq.~(81), plotted
schematically as a function of reduced doping level. Solid
(dashed) line corresponds to the value of the effective coupling
constant $w_K=2$ ($w_K=4/3$).}
\end{figure}
\begin{figure}
\caption{\label{fig:epsart} Condensation energy $\varepsilon_c$
(solid line), Eq.~(81), calculated in the case of weak
ferromagnetic ordering, SC energy gap parameter $\delta_-$ (dashed
line), Eq.~(92), plotted schematically as functions of reduced
doping level. The effective coupling constant $w_K=1$.}
\end{figure}

\end{document}